\documentclass[aps,superscriptaddress,amsmath,amssymb,twocolumn,showpacs,floatfix,reprint]{revtex4-2}
\usepackage{bm}
\usepackage[colorlinks=true, urlcolor=blue, linkcolor=blue, citecolor=blue, pdftex]{hyperref}
\usepackage{float}
\usepackage[usenames,dvipsnames]{xcolor}
\usepackage{comment}
\usepackage[utf8]{inputenc}
\usepackage{braket}
\usepackage{graphicx}

\definecolor{C0}{HTML}{1f77b4}
\definecolor{C1}{HTML}{ff7f0e}
\definecolor{C2}{HTML}{2ca02c}
\definecolor{C3}{HTML}{d62728}
\definecolor{C4}{HTML}{9467bd}
\definecolor{C5}{HTML}{8c564b}

\usepackage{todonotes}

\begin{document}

\title{Fine-tuning neural network quantum states}

\author{Riccardo Rende}
\email{rrende@sissa.it}
\affiliation{International School for Advanced Studies (SISSA), Via Bonomea 265, I-34136 Trieste, Italy}
\author{Sebastian Goldt}
\email{sgoldt@sissa.it}
\affiliation{International School for Advanced Studies (SISSA), Via Bonomea 265, I-34136 Trieste, Italy}
\author{Federico Becca}
\affiliation{Dipartimento di Fisica, Universit\`a di Trieste, Strada Costiera 11, I-34151 Trieste, Italy}
\author{Luciano Loris Viteritti}
\email{lucianoloris.viteritti@phd.units.it}
\affiliation{Dipartimento di Fisica, Universit\`a di Trieste, Strada Costiera 11, I-34151 Trieste, Italy}

\date{\today}

\begin{abstract}
Recent progress in the design and optimization of neural-network quantum states (NQSs) has made them an effective method to investigate ground-state properties of quantum many-body systems. In contrast to the standard approach of training a separate NQS from scratch at every point of the phase diagram, we demonstrate that the optimization of a NQS at a highly expressive point of the phase diagram (i.e., close to a phase transition) yieldsfeatures that can be reused to accurately describe a wide region across the transition. We demonstrate the feasibility of our approach on different systems in one and two dimensions by initially pretraining a NQS at a given point of the phase diagram, followed by fine-tuning only the output layer for all other points. Notably, the computational cost of the fine-tuning step is very low compared to the pretraining stage. We argue that the reduced cost of this paradigm has significant potential to advance the exploration of strongly-correlated systems using NQS, mirroring the success of fine-tuning in machine learning and natural language processing.
\end{abstract}

\maketitle

\section{Introduction}

Over the last decade, neural networks have emerged as the most important general-purpose machine learning tool~\cite{bengio2014, lecun2015deep}. Their versatility is evident in the most recent generation of neural networks based on the Transformer architecture~\cite{vaswani2017}, which was initially designed for natural language processing, and is now achieving state-of-the-art performance in fields as diverse as text generation~\cite{radford2019language,gpt3}, computer vision~\cite{dosovitskiy2021}, and protein contact prediction~\cite{jumper2021highly}. The success of deep neural networks is generally attributed to their ability to learn relevant features directly from data~\cite{bengio2014,lecun2015deep}. This approach replaces the classical one, where one first designs a set of well-suited features to describe the inputs, and then trains a simple machine learning algorithm (e.g., linear regression) to perform a given task. However, hand-crafting suitable features is feasible only for simple problems and becomes unfeasible when dealing with complicated tasks. By contrast, deep-learning methods learn good representations to solve the target task directly from raw data, outperforming hand-crafted representations in a variety of domains. 

In the last few years, neural networks are also increasingly used in condensed matter physics to approximate low-energy states of many-body quantum systems, for both spin and fermionic models~\cite{carleo2017,luo2019,nomuraimada2021,robledomoreno2022,roth2023, lange2024architectures,lange2023neural,pfau2020,kim2023,pescia2022,moreno2022,luo2019}.
In this context, neural-network quantum states (NQSs) parametrize the amplitude of a variational state $\ket{\Psi_{\theta}}$ expandend in a proper basis $\{\ket{\boldsymbol{\sigma}}\}$, mapping input physical configurations $\boldsymbol{\sigma}$ to complex numbers $\braket{\boldsymbol{\sigma}|\Psi_{\theta}}=\Psi(\boldsymbol{\sigma};{\theta})$. Within the variational Monte Carlo framework, the parameters ${\theta}$ of the state are optimized to minimize the variational energy $E_{\theta}=\braket{\Psi_{\theta}|\hat{H}|\Psi_{\theta}}/\braket{\Psi_{\theta}|\Psi_{\theta}}$~\cite{becca2017}.

Recently, a new parametrization of NQS which explicitly leverages the feature learning perspective has been introduced in Ref.~\cite{viteritti2023}. In this viewpoint, a deep neural network is used to construct a map from the space of the physical configurations to abstract representations in a feature space, where the determination of the low-energy properties of the systems is simplified, and then a simple shallow network transform these representations into complex numbers. The resulting architecture achieved state-of-the-art performance on one of the most famous benchmark problems of the field~\cite{rende2023}.

In this work, we address a key conceptual question that this approach raises: given a system that exhibits a phase transition, \textit{do the representations that have learned to approximate the ground state near the transition point generalize to other points of the phase diagram?} This question holds significance not only from a theoretical perspective but also from a practical one, as it provides a concrete advantage of avoiding the need of optimizing the wave function from scratch for each point in the phase diagram.

\section{Methods}
\begin{figure}[t]
  \center
  \includegraphics[width=\columnwidth]{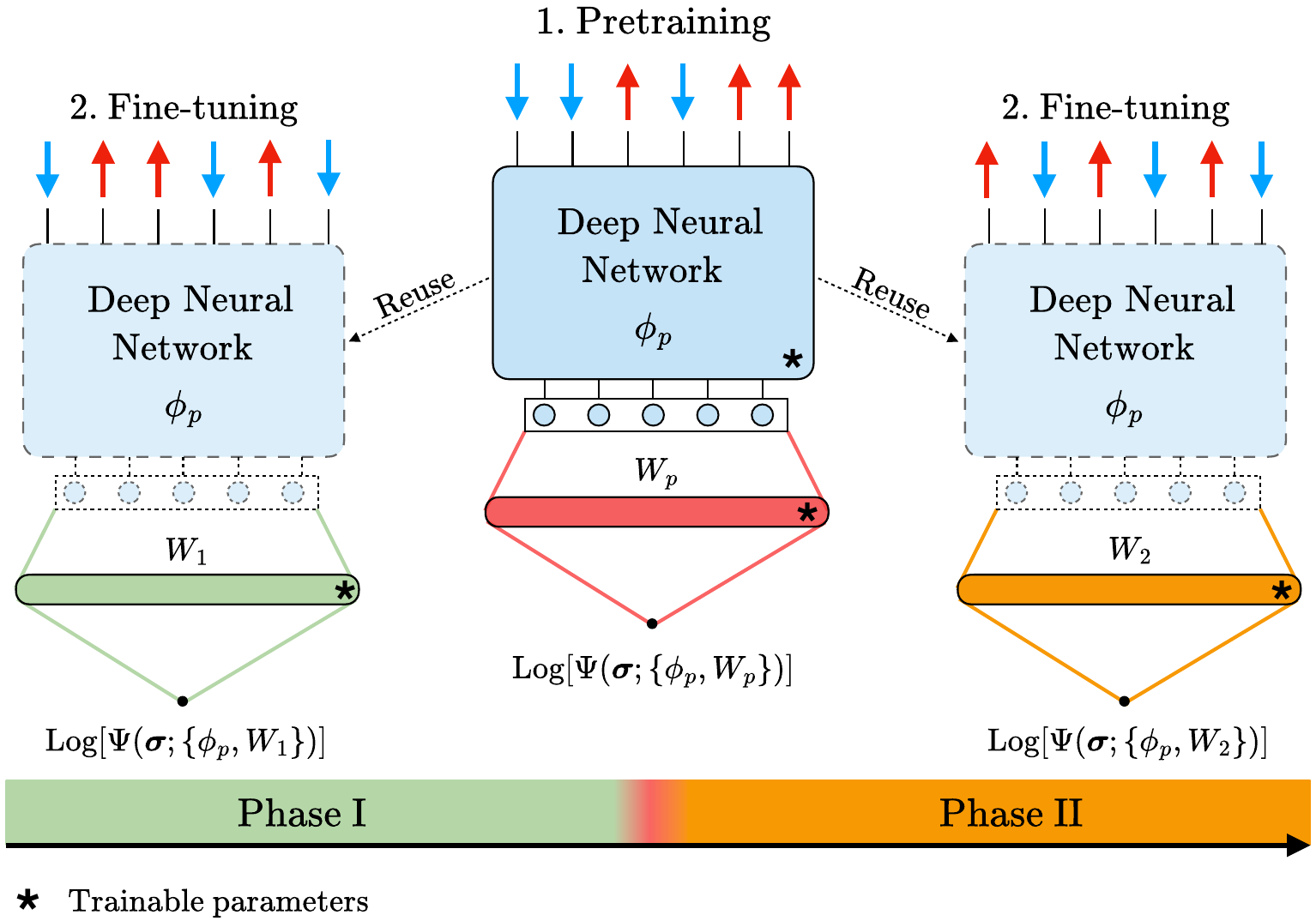}
  \caption{\label{fig:pre_training} Graphical representation of the pretraining and fine-tuning procedures. Initially, during the pretraining, the entire architecture is trained in proximity to the transition point of a given system, yielding a set of parameters $\theta_p = \{\phi_p, W_p\}$. Subsequently, in the fine-tuning stage, the parameters of the deep neural network $\phi_p$ are fixed, while the optimization process focuses exclusively on the weights of the shallow network $W$ at various points across the phase diagram.}
\end{figure}
Mirroring the feature learning perspective, we represent the NQS as the composition of two functions~\cite{viteritti2023}:
\begin{equation}\label{eq:composition}
\begin{aligned}
    \boldsymbol{z} &= V(\boldsymbol{\sigma}; \phi) \ , \\
    \text{Log}[\Psi(\boldsymbol{\sigma}; \theta)] &= f \left( \boldsymbol{z}; W \right)\ ,
\end{aligned}
\end{equation}
where we have partitioned the variational parameters into two blocks $\theta = \{\phi, W\}$. The function $V(\boldsymbol{\sigma}; \phi)$ is parameterized through a \textit{deep} neural network, mapping physical configurations $\boldsymbol{\sigma}$ into a feature space, thereby generating for each input configuration $\boldsymbol{\sigma}$ a hidden representation $\boldsymbol{z}(\boldsymbol{\sigma}) \in \mathbb{R}^d$, with $d$ the dimension of the feature space (an adjustable hyperparameter of the network). Conversely, $f(\boldsymbol{z}; W)$ is a \textit{shallow} fully connected neural network used to generate a single scalar value $f(\boldsymbol{z}; W) \in \mathbb{C}$ from hidden representations $\boldsymbol{z}$. 

Given a system undergoing a phase transition, we want to investigate whether the representations learned near the transition point may be generalized to other points of the phase diagram. We perform the following experiment in two steps, as illustrated in Fig.~\ref{fig:pre_training}:
\begin{figure*}[t]
  \center
  \includegraphics[width=\linewidth]{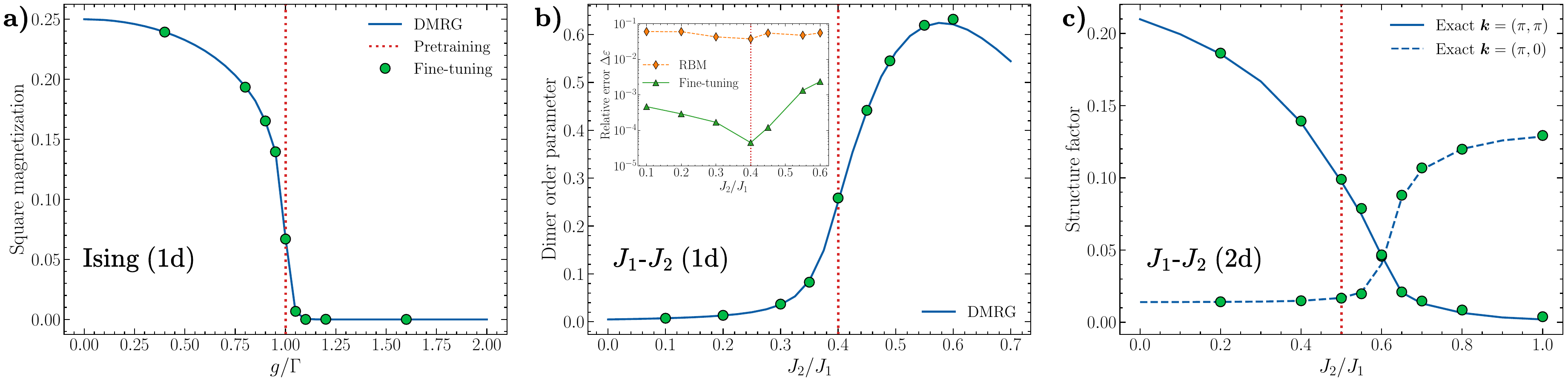}
  \caption{\label{fig:fig_orderparams} (a) Ising chain. A ViT with hyperparameters $h=12$, $d=72$, $n_l=4$ is pretrained at $g/\Gamma=1$, on a chain with $N=100$ sites. After the fine-tuning, the square magnetization order parameter is computed and compared to DMRG results (bond dimension $\chi = 10^3$). (b) Heisenberg $J_1$-$J_2$ chain. A ViT with hyperparameters $h=12$, $d=192$, $n_l=4$ is pretrained at $J_2/J_1=0.4$, on a chain with $N=100$ sites. After the fine-tuning, the dimer order parameter is computed and compared to DMRG results ($\chi = 10^3$). Inset : Relative error $\Delta \varepsilon$ (with respect to DMRG) of the same fully connected network (RBM) trained on the hidden representations generated by the pretrained ViT and directly on configurations. (c) Two dimensional Heisenberg $J_1$-$J_2$. A ViT with hyperparameters $h=18$, $d=216$, $n_l=8$ is pretrained at $J_2/J_1=0.5$, on a $6\times6$ square lattice. After the fine-tuning, the structure factors at $\boldsymbol{k}=(\pi,\pi)$ and $\boldsymbol{k}=(0,\pi)$ are computed and compared to exact diagonalization results.}
\end{figure*}
\begin{enumerate}
    \item We \textit{pretrain} the entire network, optimizing it to approximate the ground state at a single point of the phase diagram situated in the vicinity of the phase transition. The pretraining stage yields a set of optimized parameters $\{\phi_{p}, W_{p}\}$.
    \item Using the features constructed by the deep network (thus fixing its variational parameters ${\phi}_{p}$) we \textit{fine-tune} the model by optimizing only the parameters $W$ of the output layer to approximate the ground states in the other points of the phase diagram, on both sides of the phase transition.
\end{enumerate}

The pretraining of the architecture is carried out near the critical point, where long-range correlations are present and may be established in the body of the NQS. Then, the last (shallow) output layer, which is fine-tuned in a different point in the phase diagram, can either reinforce correlations and establish true long-range order or weaken them and yield to a short-range state (or even keep the state critical). On the other hand, in trivial phases, where only a few configurations have non-zero amplitudes, the ability of pretrained networks to generalize away from these phases is limited (see Appendix~\ref{sec:starting}).

We apply this procedure on finite systems and measure physical properties (e.g., order parameters) of various systems exhibiting, in the thermodynamic limit, phase transitions of different nature. In all cases, the features extracted during the pretraining stage, close to transition points, lead to excellent results after fine-tuning at the other points of the phase diagram~\cite{tang2024learning}. This result reflects how neural networks can capture the essential quantum fluctuations in the vicinity of a phase transition. We stress that this approach differs from the standard transfer-learning paradigm in which a neural network is initially trained to solve a specific task, and then all of the parameters are trained to solve a different task. To the best of our knowledge, fine-tuning experiments on NQS have not been explored previously.

The methodology outlined here is generally applicable to any deep neural network, but to be concrete, in the following we parametrize the function $V(\boldsymbol{\sigma}; \phi)$ using a Vision Transformer (ViT) with real-valued parameters~\cite{viteritti2023,rende2023}. This choice is suggested by the flexibility of the Transformer architecture~\cite{vaswani2017,dosovitskiy2021}, already used to achieve highly accurate results in various types of systems, both one- and two-dimensional~\cite{melko2024language,czischek2023,rende2023,viteritti2023,luo2022, diluo2023,viteritti20231d}. The hyperparameters of this architecture are the number of heads $h$ of the Multi-Head Factored Attention Mechanism~\cite{rende2023b}, the embedding dimension $d$ and the number of layers $n_l$ (a detailed description of the architecture is reported in Ref.~\cite{viteritti2023}). In addition, the function $f(\boldsymbol{z}, W)$ is defined as
\begin{equation}\label{eq:rbm}
    f(\boldsymbol{z}; W) = \sum_{\alpha=1}^K  g( b_{\alpha} + \boldsymbol{w}_{\alpha} \cdot \boldsymbol{z} ) \ ,
\end{equation}
where the number of neurons $K$ is chosen to be equal to~$d$ and $2 \times d$ in the pretraining and in the fine-tuning steps, respectively. In order to describe nonpositive ground states, the parameters $W=\{b_{\alpha}, \boldsymbol{w}_{\alpha}\}_{\alpha=1}^{K}$ of the linear transformation in Eq.~\eqref{eq:rbm} are taken to be complex valued. Here, we set $g(\cdot) = \log\cosh(\cdot)$, thus $f(\boldsymbol{z}, W)$ represents the well-known Restricted Boltzmann Machine (RBM) introduced by~\citet{carleo2017}. The crucial difference is that, in this case, it is not applied directly on physical configurations $\boldsymbol{\sigma}$, but instead on hidden representations $\boldsymbol{z}$ generated by the Transformer~\cite{viteritti2023}. We remark that this framework offers a huge computational advantage, since it requires the costly optimization of the full architecture, including the feature extractor $V(\boldsymbol{\sigma}; \phi)$, only once in the pretraining step. Then, with the addition of a minimal cost, the targeted optimization of the RBM can be used to obtain an accurate description of the physical properties of the system in a wide region across the transition point. In what follows, we focus on spin $S=1/2$ models on a lattice considering system sizes where numerically exact solutions are available for comparison. In this case $\Psi(\boldsymbol{\sigma};{\theta})$ refers to the amplitude of the variational state $\ket{\Psi_{\theta}}$ in the computational basis having a definite spin value in the $z$ direction, i.e., $\{\ket{\boldsymbol{\sigma}} = \ket{\sigma_1^z, \cdots, \sigma_N^z} \}$ with $\sigma_i^z = \pm 1$.

Finally, we mention a few details on the Monte Carlo procedure adopted here. During the pretraining stage, we perform $N_{\text{opt}}=10000$ optimization steps. Then, during the fine-tuning stage, the number of steps is reduced to $N_{\text{opt}} = 3000$. For all the simulations, we estimate stochastically the observables choosing a number of samples of $M=3000$. The optimization of the variational parameters is performed with the Stochastic Reconfiguration (SR) method~\cite{sorella2005}. In particular, working with variational states featuring approximately $P=10^6$ parameters, we employ the alternative formulation of SR~\cite{rende2023,chen2023} efficient in the regime $P \gg M$ (available in NetKet~\cite{netket3} under the name of \href{https://netket.readthedocs.io/en/latest/api/_generated/experimental/driver/netket.experimental.driver.VMC_SRt.html#netket.experimental.driver.VMC_SRt}{\texttt{VMC\_SRt}}.). We use a cosine decay scheduler for the learning rate, setting the initial value to $\tau=0.03$.

\section{Results}

\subsection{The one-dimensional quantum Ising model}

We start by considering the one-dimensional Ising model in transverse magnetic field, described by the following Hamiltonian
\begin{equation}\label{eq:is_ham}
    \hat{H} = -\Gamma\sum_{i=1}^N \hat{S}_i^z \hat{S}_{i+1}^z - g \sum_{i=1}^N \hat{S}_i^x \ ,
\end{equation}
where $\hat{S}_i^x$ and $\hat{S}_i^z$ are spin-$1/2$ operators on site $i$. The ground-state wave function, for $g \ge0$, is positive definite in the computational basis. 

In the thermodynamic limit, the ground state exhibits a second-order phase transition at $g/\Gamma=1$, from a ferromagnetic ($g/\Gamma<1$) to a paramagnetic ($g/\Gamma>1$) phase. On finite systems with $N$ sites, the estimation of the critical point can be obtained from the long-range behavior of the spin-spin correlations, i.e., $m^2(r) = 1/N \sum_{i=1}^N \langle \hat{S}_i^z \hat{S}_{i+r}^z \rangle$ (specifically, we can consider the largest distance $r=N/2$, which gives the square magnetization). 

First, we pretrain the full architecture at the critical point $g/\Gamma=1$. Then, we fine-tune only the output layer at different values of the external field, from $g/\Gamma=0.4$ to $g/\Gamma=1.6$, i.e., in both ferromagnetic and paramagnetic phases. The results for $N=100$ with periodic-boundary conditions are shown in Fig.~\ref{fig:fig_orderparams}(a), compared with density-matrix renormalization group (DMRG)~\cite{white1992} calculations (on the same system). The high level of accuracy demonstrates that the fine-tuned network is effective in the prediction of the order parameter. Remarkably, the fine-tuning procedure involves optimizing merely $6.6\%$ of the total parameters, which is ten times faster than optimizing the entire network and demands significantly less GPU memory (see Appendix~\ref{sec:mem} for a detailed description of the GPU memory requirements). The remarkable fact is that, by exclusively adjusting the parameters of the output (fully connected) layer and keeping the clusters of the hidden representation fixed, it is possible to effectively describe both ordered and disordered phases.
\begin{figure*}[t]
  \includegraphics[width=0.95\linewidth]{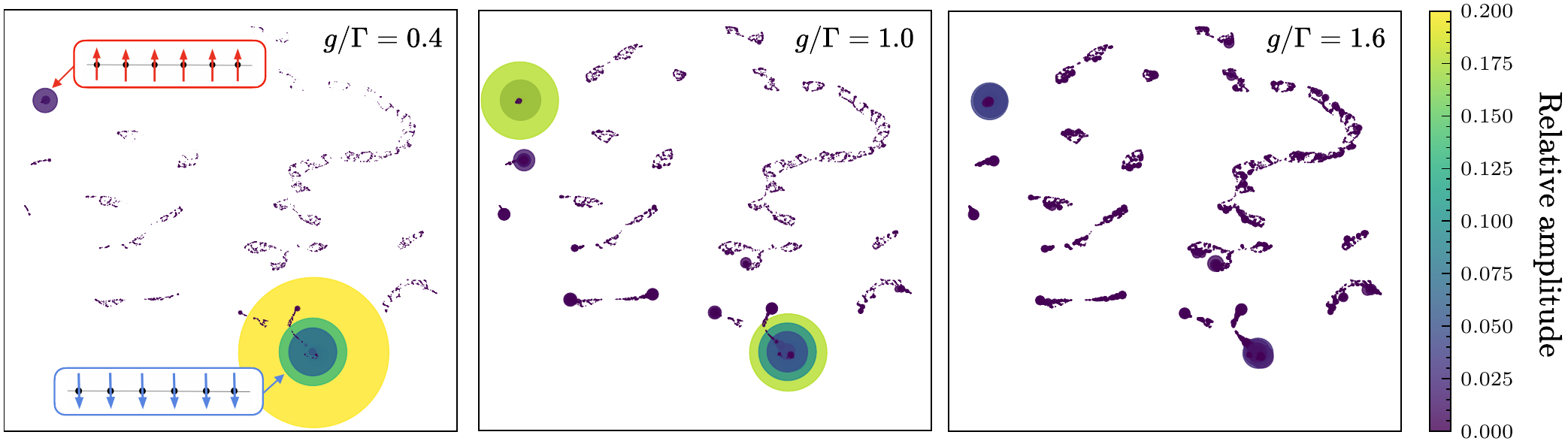}
  \caption{\label{fig:umap_amps}Dimensional reduction of the hidden representations for a set of $M=3000$ configurations built using a ViT pretrained at $g/\Gamma=1$ with hyperparameters $h=12$, $d=72$, and $n_l=4$ for a system of $N=100$ sites.
  The data points represent UMAP projections of vectors $\boldsymbol{z}$. Both the colors and sizes of the points are related to the amplitudes predicted after the fine-tuning procedure at three distinct points along the phase diagram: ordered phase $g/\Gamma=0.4$ (left panel), transition point $g/\Gamma=1$ (central panel) and disordered phase $g/\Gamma=1.6$ (right panel).}
\end{figure*}

In the following, we want to gain insights into the learning process of the fine-tuning stage. For that, we sample a set of $M$ configurations $\{\boldsymbol{\sigma}_i\} \sim |\Psi(\boldsymbol{\sigma};\theta_p)|^2$ from the pretrained network and show the corresponding amplitudes after the finetuning procedure [visualizing them on top of UMAP
\footnote{UMAP~\cite{mcinnes2020} is a general purpose dimension reduction technique for machine learning. It is constructed from a theoretical framework based in Riemannian geometry and algebraic topology, see Ref.~\cite{mcinnes2020} for more details.}
projections of the hidden representations $\boldsymbol{z}_p(\boldsymbol{\sigma}_i)$, for $i=1, \dots, M$]; see Fig.~\ref{fig:umap_amps}. To highlight the differences, both color and size of each point are proportional to their amplitudes. At the transition point ($g/\Gamma=1$), the configurations with all parallel spins (either up or down along $z$) have the largest amplitude; other configurations, with a few spin flips have also considerable weights (see middle panel). In the ordered phase ($g/\Gamma=0.4$), only one of these fully polarized configurations is ``selected'', i.e., frequently visited along the Monte Carlo sampling, and the amplitudes for all other configurations are practically negligible (left panel). This effect is related to the difficulty of simple sampling approaches (that performs local spin flips) to overcome the (large) barrier that separates the two ground states, which are almost degenerate on finite systems. By contrast, in the disordered phase ($g/\Gamma=1.6$), many configurations have similar amplitudes: the two fully polarized configurations showing a reduced weight compared to all the others (right panel). A brief discussion about the connection between the features extracted by the NQS and the order parameter is given in Appendix~\ref{sec:feature}.

\subsection{The one-dimensional $J_1$-$J_2$ Heisenberg model}

In order to assess the accuracy of our method on more complicated systems, specifically with non-positive ground states in the computational basis, we investigate the frustrated $J_1$-$J_2$ Heisenberg model:
\begin{equation}\label{eq:J1J2_ham}
  \hat{H} = J_1\sum_{\langle i,j \rangle} \hat{\boldsymbol{S}}_{i}\cdot\hat{\boldsymbol{S}}_{j} + J_2\sum_{\langle\langle i,j \rangle\rangle} \hat{\boldsymbol{S}}_{i}\cdot\hat{\boldsymbol{S}}_{j} \ ,
\end{equation}
where $\hat{\boldsymbol{S}}_{i}=(S_i^x,S_i^y,S_i^z)$ and $J_1>0$ and $J_2 \ge 0$ are antiferromagnetic couplings for nearest- and next-nearest neighbors, respectively. 

We first discuss the results in one dimension. Here, the ground-state phase diagram shows two phases, separated by a Berezinskii-Kosterlitz-Thouless transition at $(J_2/J_1)_c = 0.24116(7)$~\cite{eggert1996}: a gapless phase with no order whatsoever and a gapped one, with long-range dimer order. On finite systems, the latter one may be extracted from the long-distance behavior of the dimer-dimer correlation functions ${D(r) = \braket{\hat{S}^z_1\hat{S}^z_2\hat{S}^z_r\hat{S}^z_{r+1}} - \braket{\hat{S}^z_1\hat{S}^z_2}\braket{\hat{S}^z_r\hat{S}^z_{r+1}}}$~\cite{lacroix2011book,capriotti2003}. Specifically, performing a finite-size scaling, an estimation of the dimer order parameter can be obtained as $D^2 = 9 |D(N/2-1) - 2 D(N/2) + D(N/2+1)|$~\cite{lacroix2011book,capriotti2003}. However, we emphasize that the order parameter is exponentially small close to the transition, making it difficult to extract an accurate estimation of the actual value of $(J_2/J_1)_c$ (indeed, the location of the transition may be easily obtained by looking at the level crossing between the lowest-energy triplet and singlet excitations~\cite{eggert1996}). As before, we pretrain at a given point, here $J_2/J_1=0.4$, and optimize the output layer of the network for different values of the frustrating ratio, in both the gapless and gapped regions. The results for $N=100$ (with periodic boundary conditions) are reported in Fig.~\ref{fig:fig_orderparams}b), again compared to DMRG calculations on the same system. In addition, in the inset of Fig.~\ref{fig:fig_orderparams}b), we compare the relative energy error $\Delta\varepsilon$ (with respect to the DMRG energies) of an RBM trained directly on the physical configurations~\cite{carleo2017} and of a fine-tuned ViT. This analysis underscores the importance of exploiting the features constructed by the pretrained ViT, resulting in an accuracy gain of more than two orders of magnitude with respect to the same network trained directly on configurations.

\begin{table}[b]
    \centering
    \begin{tabular}{p{0.15\linewidth}p{0.25\linewidth}p{0.25\linewidth}c}
        \hline\hline
         \centering $\boldsymbol{J_2/J_1}$&  \centering \textbf{DMRG}&  \centering \textbf{ViT}& \textbf{Fine-tuning}\\ \hline
          \centering0.10&   \centering-0.425417395&   \centering-0.4254174&  -0.425218\\ 
          \centering0.20&   \centering-0.408572967&  \centering-0.4085728& -0.408453\\ 
          \centering0.30&   \centering-0.393126745&  \centering-0.3931204& -0.393059\\ 
          \centering0.40&  \centering-0.380387370&  \centering-0.3803726& -0.380370\\ 
         \centering0.60&  \centering-0.380804138&  \centering-0.3807913& -0.379902\\
         \hline
    \end{tabular}
    \caption{Variational ground-state energies for the $J_1$-$J_2$ Heisenberg chain with system size $N=100$. The DMRG computations are conducted employing a bond dimension up to $\chi = 10^3$ under periodic boundary conditions. For both instances involving ViT, one trained from scratch and another pretrained at $J_2/J_1=0.4$ followed by fine-tuning, the Monte Carlo error attributed to finite sampling effects affects the last digit of the reported results.}
    \label{tab:energies}
\end{table}
For completeness, we report the ground-state energies for various frustration ratios $J_2/J_1$ in Table~\ref{tab:energies}. They are obtained through three distinct methodologies: DMRG, ViT trained from scratch, and ViT pretrained at ${J_2/J_1=0.4}$ and subsequently fine-tuned for other frustration ratios. Notably, the fine-tuned ViT exhibits remarkable accuracy when compared to DMRG results, reaching an accuracy $\Delta \varepsilon \lesssim 10^{-3}$ for all the values of the frustration ratio in the interval $J_2/J_1 \in [0.1, 0.6]$.
\begin{figure}[t]
  \center
  \includegraphics[width=0.9\columnwidth]{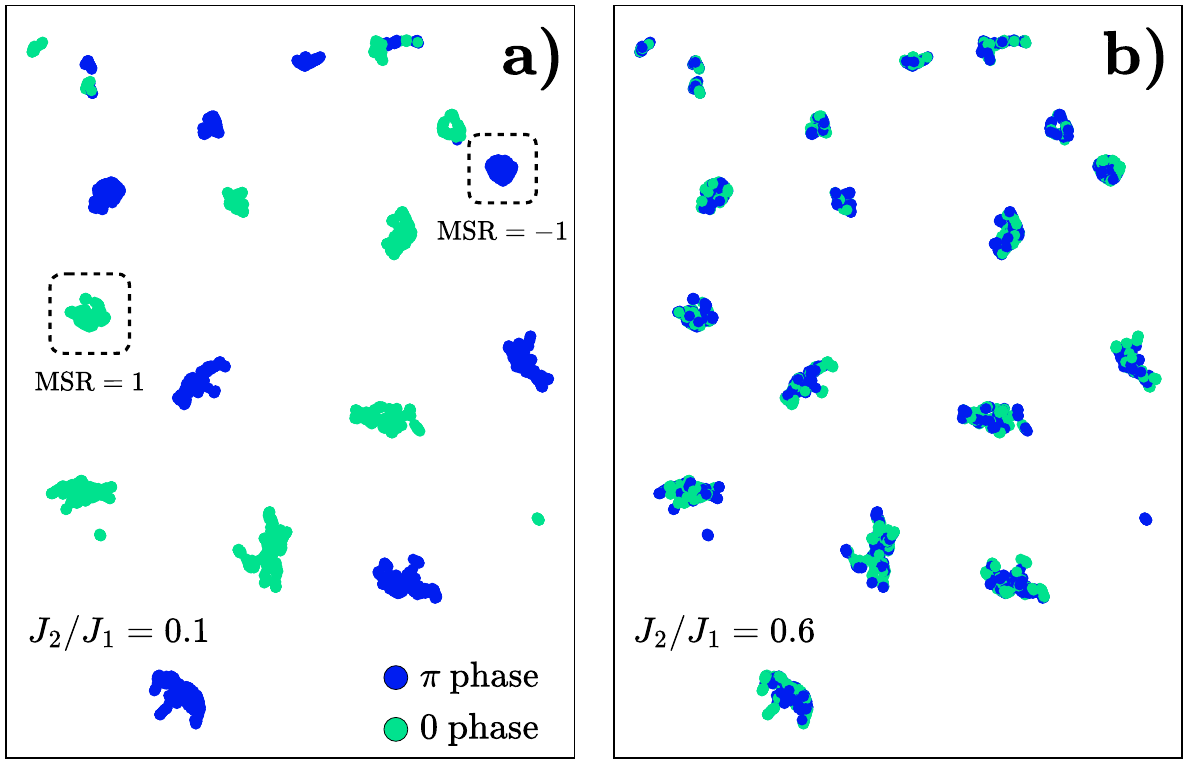}
  \caption{\label{fig:umap_signs} Graphical representation of the hidden representations for the $J_1$-$J_2$ Heisenberg chain. The data points, corresponding to a sample of $M=3000$ physical configurations, represent UMAP projections of vectors $\boldsymbol{z}_p$ generated by a ViT with hyperparameters $h=12$, $d=192$, and $n_l=4$, pretrained at the point $J_2/J_1=0.4$ for a system size of $N=100$. The depicted colors correspond to the predicted phases ($0$ or $\pi$) after fine-tuning at two specific points within the phase diagram: $J_2/J_1=0.1$ [panel (a)] and $J_2/J_1=0.6$ [panel (b)]. Panel (a) reveals a close resemblance between the cluster structure identified during the pretraining at $J_2/J_1=0.4$, which matches the Marshall Sign Rule.}
\end{figure}
Let us move to the discussion of how the output layer can modify the sign structure during the fine-tuning step. For the $J_1$-$J_2$ Heisenberg chain, the sign structure of the ground-state wave function is not known except for $J_2=0$, where the so-called Marshall Sign Rule (MSR)~\cite{marshall1955} applies. However, even for large system sizes, the MSR constitutes an accurate approximation of the sign structure up to $J_2/J_1 \le 0.5$~\cite{viteritti2022}. In Fig.~\ref{fig:umap_signs}, we show the predicted phases ($0$ or $\pi$), on top of the UMAP projections of the vectors $\boldsymbol{z}_p$ generated by the pretrained network at $J_2/J_1=0.4$. At $J_2/J_1=0.1$ [see Fig.~\ref{fig:umap_signs}(a)], after the fine-tuning procedure, the signs exactly match the ones obtained at $J_2/J_1=0.4$ (not shown). This is because, at the pretraining point, where the clusters are formed, the MSR remains a highly accurate approximation of the ground-state sign structure.
By contrast, for $J_2/J_1=0.6$, this is no longer true, and the output layer must adjust the phases accordingly [see Fig.~\ref{fig:umap_signs}(b)]; still, the fine-tuned ViT performs better than a RBM trained on spin configurations; see Appendix~\ref{sec:MSR} for a detailed discussion.

\subsection{The two-dimensional $J_1$-$J_2$ Heisenberg model}

Finally, we consider the two-dimensional $J_1$-$J_2$ Heisenberg model on an $L \times L$ square lattice. The ground state of this model features magnetic order in the two limits $J_1 \ll J_2$ and $J_1 \gg J_2$. Its presence can be characterized with the spin structure factor
${S(\boldsymbol{k})= \sum_{\boldsymbol{R}}e^{i\boldsymbol{k}\cdot\boldsymbol{R}} \braket{\hat{\boldsymbol{S}}_{\boldsymbol{0}}\cdot \hat{\boldsymbol{S}}_{\boldsymbol{R}}}}$, 
where $\boldsymbol{R}$ runs over all the lattice sites of the square lattice.
Specifically, when $J_2=0$ the model reduces to the unfrustrated Heisenberg model where long-range Néel order is present~\cite{calandra1998,sandvik1997}. The latter one can be detected by measuring ${m^2_{\text{Néel}}=S(\pi,\pi)/L^2}$. In the opposite regime $J_2/J_1 \rightarrow \infty$, the system exhibits instead columnar magnetic order, identified by the order parameter ${m^2_{\text{stripe}}=[S(0,\pi) + S(\pi,0)]/(2 L^2)}$. In the intermediate region, around $J_2/J_1 \approx 0.5$ the system is highly frustrated and the nature of the ground state is still under debate~\cite{nomuraimada2021,BeccaGutz2013,ferrari2020,wang2022,gong2014}. Here, we limit ourselves to the $6 \times 6$ system, where exact diagonalizations are possible (no DMRG calculations on the structure factor are available on larger systems with periodic boundary conditions). We first perform the pretraining at $J_1/J_2 = 0.5$, then perform the fine-tuning for $0.2<J_2/J_1<1$ and evaluate the order parameters $m^2_{\text{Néel}}$ and $m^2_{\text{stripe}}$, see Fig.~\ref{fig:fig_orderparams}c). Remarkably, even for this complicated two-dimensional model, the correct behavior of the two magnetic order parameters can be reconstructed with great accuracy starting from a single pretrained ViT.

\section{Conclusions}

We showed that, for several physical systems exhibiting phase transitions, pretraining a Neural-Network Quantum State near the transition point yields a set of features that can be fine-tuned to obtain accurate descriptions of the phase diagram. In addition, the analysis of the feature space facilitated the extraction of valuable insights into the structure of the wave function. Furthermore, the fine-tuning process proves to be computationally more efficient, in terms of both time and memory, compared to traditional approaches in which the full training is required at each point of the phase diagram. We contend that, akin to the prevalent approach in machine learning~\cite{wolf2020}, the employment of pretrained networks holds great potential for advancing the exploration of physical systems through NQS. This aligns with the findings of a recent study by Ref.~\cite{scherbela2024towards}, demonstrating that, in electronic-structure problems, a single wave function can be employed to investigate multiple compounds and geometries. A similar framework has also been proposed for decoding quantum error-correcting codes with generative modeling~\cite{cao2023qecgpt}. Exploring extensions of this approach to perform fine-tuning across different physical models stands as a crucial topic for future studies, as well as the development of techniques to automatically identify the most expressive pretraining point~\cite{santos2021,vannieuwenburg2017, zen2020,lewis2023}. Another possible application could focus on approximating the real-time dynamics~\cite{SchmittDyn2020,HeylDyn2023} by adjusting only the parameters of the shallow output network, thereby solving the problem within the feature space rather than the configuration space. It would be intriguing to further explore the precise physical meaning of the learned features, such as establishing a connection between the clusters in the feature space and the order parameters. \\

\begin{acknowledgments}
We thank A. Laio, G. Carleo, F.~Vicentini and M.~Imada for useful discussions. We acknowledge the CINECA award under the ISCRA initiative, for the availability of high-performance computing resources and support. The DMRG calculations
have been performed within the ITensor library~\cite{itensor}.
\end{acknowledgments}

\appendix

\section{Choosing different pretraining points}\label{sec:starting}
\begin{figure}[t]
  \center
  \includegraphics[width=0.9\columnwidth]{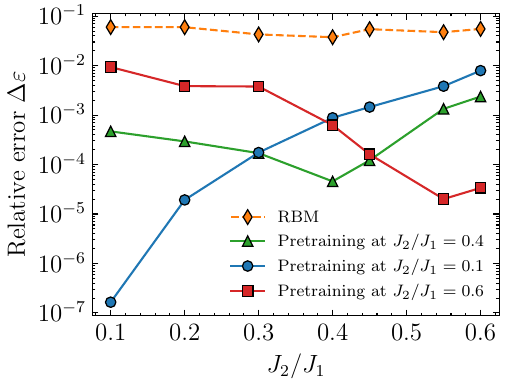}
  \caption{\label{fig:accuracy} Relative error $\Delta \varepsilon$ of the energy with respect to DMRG for the $J_1$-$J_2$ Heisenberg model on a chain [see Eq.~\eqref{eq:J1J2_ham}] of $N=100$ sites. The curves are obtained performing the fine-tuning procedure starting from different pretraining points generated by a ViT with hyperparameters $h=12$, $d=192$, $n_l=4$. Specifically, we set $J_2/J_1=0.4$ (green triangles), $J_2/J_1=0.1$ (blue circles), and $J_2/J_1 = 0.6$ (red squares). The accuracy of the same fully connected network (RBM) optimized on the physical configurations is also reported for comparison (orange diamonds).}
\end{figure}

The accuracy of the fine-tuning across various points on the phase diagram is influenced by the choice of the pretraining point. In our study, we have always pretrained near transition points, where we expect better generalization properties as discussed in the main text. Here, we investigate how the accuracy of the fine-tuned results varies when choosing different pretraining points, for example within the bulk of one phase. In Fig.~\ref{fig:accuracy}, we show the accuracy of the energy $\Delta \varepsilon$ relative to DMRG calculations for the $J_1$-$J_2$ Heisenberg model on a chain [see Eq.~\eqref{eq:J1J2_ham}] of $N=100$ sites (assuming periodic boundary conditions).
The transition point of the model in the thermodynamic limit is $(J_2/J_1)_c=0.24116(7)$; however, on a finite system with $N=100$ sites, the point exhibiting the maximum slope in the dimer order parameter occurs around $J_2/J_1 = 0.4$ (refer to the central panel of Fig.~\ref{fig:fig_orderparams}). The accuracy of the fine-tuned energies, using $J_2/J_1=0.4$ as the pretraining point, is approximately $\Delta \varepsilon \approx 10^{-3}$ within the interval $J_2/J_1 \in [0.1, 0.6]$ (green triangles in Fig.~\ref{fig:accuracy}). Conversely, pretraining from $J_2/J_1=0.1$ (blue circles in Fig.~\ref{fig:accuracy}) yields higher accuracy before $J_2/J_1=0.4$, but as the distance from the pretraining point increases the accuracy deteriorates to approximately $\Delta \varepsilon \approx 10^{-2}$. A similar behavior can be observed when choosing $J_2/J_1=0.6$ as the pretraining point (red squares in Fig.~\ref{fig:accuracy}). 

It is interesting to note that the accuracy of the network pretrained at $J_2/J_1 = 0.1$ (blue circles) deteriorates by four orders of magnitude when fine-tuning at $J_2/J_1 = 0.4$. In contrast, the network pretrained at $J_2/J_1 = 0.4$ (green triangles) loses less than one order of magnitude in accuracy when finetuning at $J_2/J_1 = 0.1$, with the error rising from $\Delta\epsilon \approx 10^{-4}$ to $\Delta\epsilon \approx 10^{-3}$. This result suggests that features learned near the phase transition are more robust for generalization compared to those learned within the bulk of a phase.
Consequently, selecting a pretraining point that lies near the transition appears to strike the optimal balance, yielding an accuracy roughly consistent across all other points within the phase diagram. Nevertheless, it is worth noting that training the RBM in the hidden space consistently outperforms direct training on physical configurations, as illustrated by the orange diamonds in Fig.~\ref{fig:accuracy}.
\section{Memory Efficiency in Fine-tuning and Pretraining Processes}\label{sec:mem}
The primary constraint in training neural networks with a large number of parameters arises from the restricted memory capacity of contemporary graphical processing units (GPUs), rather than their computational speed. Specifically, this limitation is associated to the back-propagation algorithm~\cite{bengio2014}, which crucial for evaluating the gradients of the network efficiently, but whose memory cost scales with the depth of the computation. 
Consider a deep neural network that takes an input vector $\boldsymbol{x}$ and produces a scalar output $f(\boldsymbol{x}, \theta) \in \mathbb{R}$, where $\theta$ is a vector of trainable parameters. For simplicity, we arrange these parameters as $\theta = \text{Concat}(\theta_0, \dots, \theta_{n_l})$, where $\theta_l$ is a vector containing all the $P_l$ parameters of the $l$-th layer, and $P$ is the total number of parameters across all layers, i.e., $P=\sum_{l=1}^{n_l}P_l$. Additionally, assume that,  when computing the output, the network generates $K$ intermediate activations $a_k$, each of size $A_k$, and $A$ is the overall number of activations calculated as $A=\sum_{k=1}^K A_k$. For a batch of $M$ distinct input vectors, the loss function can be defined as $\mathcal{L}(\theta)=(1/M) \sum_{i=1}^M L[f(\boldsymbol{x}_i, \theta)]$. To efficiently backpropagate the gradients of the loss with respect to the parameters, it is necessary to store all the $A$ activations. Thus the total memory cost of the algorithm scales with the depth of the computations and is expressed as $M\times(A + \text{max}_l P_l)$ (neglecting the cost of storing all $P$ weights). On the other hand, for the forward pass the memory cost is independent of the computation depth and is equal to $M \times (\text{max}_k A_k + \text{max}_l P_l)$. Further details can be found in Ref.~\cite{novak22}.
Notice that, during the fine-tuning process, the memory-intensive backward pass over the deep network becomes unnecessary. In the context of this paper, for the used ViT architectures, the memory needed during the fine-tuning stage is approximately ten times less than what is required during the pretraining stage. The backpropagation of gradients constitutes the primary memory bottleneck, even when employing the Stochastic Reconfiguration optimization method~\cite{rende2023}. This method requires the allocation of a matrix containing $4M^2$ real numbers, where $M$ denotes the number of samples used in optimization. With double precision, this memory requirement translates to $32M^2/10^9$ GB. In our optimizations, with $M=3000$, the memory usage is approximately 0.3 GB. This is two orders of magnitude smaller than the memory required for the backward pass during pretraining.

\begin{figure}[t]
  \center
  \includegraphics[width=1.0\columnwidth]{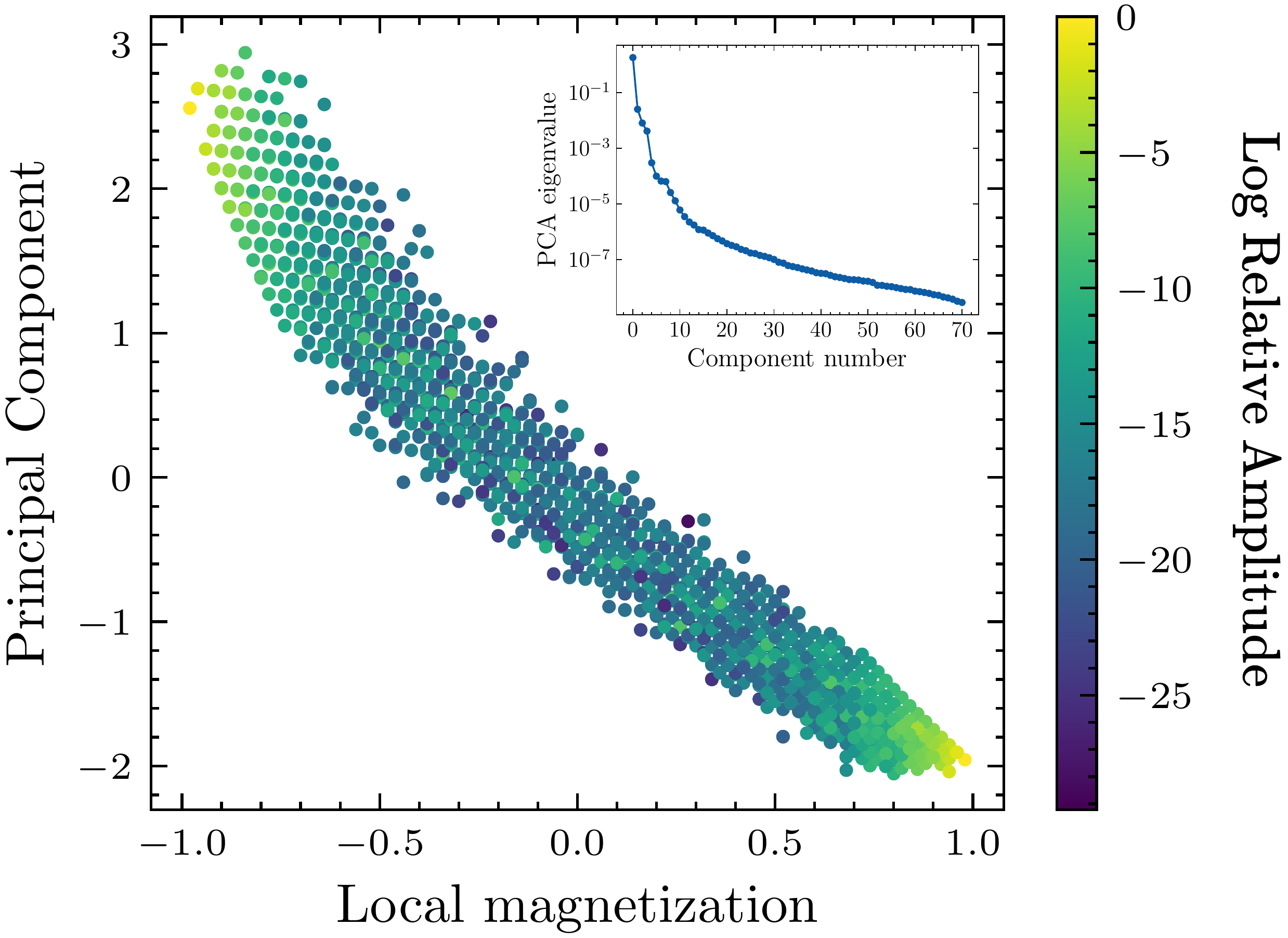}
  \caption{\label{fig:pca} Correlation between the local magnetization $\sum_{i=1}^N \sigma_i$ and the principal component of the hidden representations of the configurations associated to the ViT used to obtain the results for the Ising model in transverse field (see {\it Results} section). In the inset, the PCA spectrum is shown.}
\end{figure}

\begin{figure}[t]
  \center
  \includegraphics[width=\columnwidth]{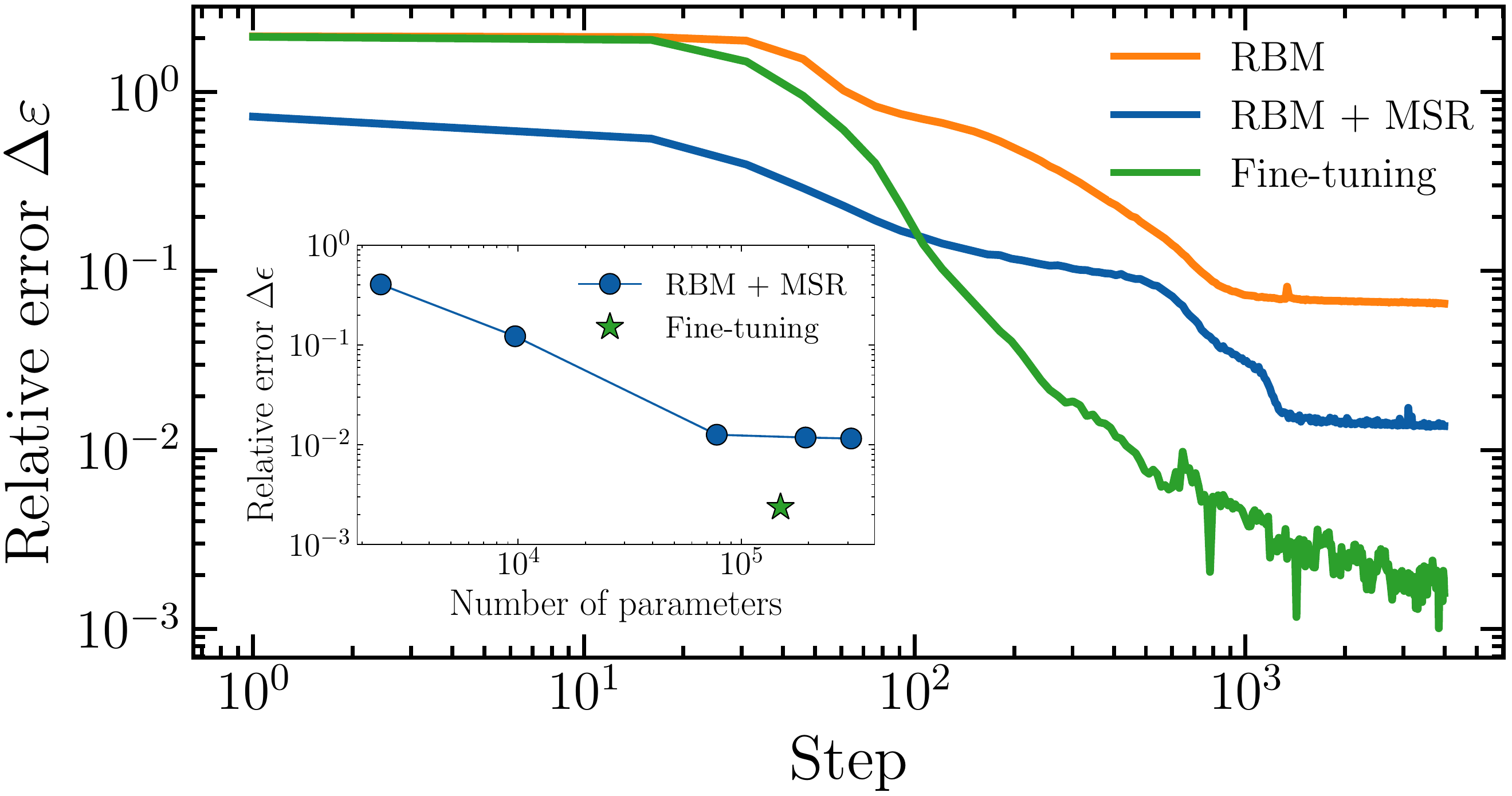}
  \caption{\label{fig:comp_MSR} Relative error in energy $\Delta \varepsilon$, compared to DMRG, plotted as a function of the optimization steps for the $J_1$-$J_2$ Heisenberg model [refer to Eq.~\eqref{eq:J1J2_ham}] with $J_2/J_1=0.6$ on a system of $N=100$ sites. The orange curve represents the variational energy obtained using a RBM with $K=384$ hidden neurons and $77568$ parameters. The blue curve depicts the same network with the addition of the Marshall Sign Rule as a prior for the sign structure. In contrast, the green curve is obtained by optimizing the same network on top of the hidden representation $\boldsymbol{z}$ generated by the Transformer with hyperparameters $h=12$, $d=192$, $n_l=4$ at $J_2/J_1=0.4$. Inset: Relative error in energy $\Delta \varepsilon$ of a RBM trained with the MSR prior as a function of the number of parameters. For comparison, the accuracy of the finetuned network is also shown.}
\end{figure}

\section{Connecting the features to order parameters}\label{sec:feature}

In this appendix we examine the Ising model in a transverse field [see Eq.~\eqref{eq:is_ham}] aiming to establish a connection between the features learned by the ViT optimized at $g/\Gamma=1$ and the magnetization order parameter that controls the phase transition. To achieve this, we consider a fixed batch of physical spin configurations \(\{\boldsymbol{\sigma}_i\} \sim |\Psi(\boldsymbol{\sigma};\theta_p)|^2\). We then compute the corresponding hidden representation and perform principal component analysis (PCA) on it. In Fig.~\ref{fig:pca}, we plot the principal component against the local magnetizations of the spin configurations, i.e., \(\sum_{i=1}^N \sigma_i\). The two quantities exhibit a strong correlation.
This organization of configurations in the feature space simplifies the description of the physics of the system, allowing for an easy transition from the ordered phase to the disordered phase. It would be interesting to extend this analysis to other systems, where the order parameter is associated to an off-diagonal operator in the computational basis.

\section{Comparison with Marshall Sign Prior}\label{sec:MSR}

A relevant question is trying to understand which kind of prior information is encoded in the features generated by the pretrained network.
We focus on the $J_1$-$J_2$ Heisenberg model on a chain [see Eq.~\eqref{eq:J1J2_ham}], with $N=100$ sites, fixing the value of the frustration ratio to $J_2/J_1=0.6$. We consider a Restricted Boltzmann Machine (RBM) with $K$ hidden neurons [see Eq.~\eqref{eq:rbm}]. This network is employed in two distinct manners: trained directly on the physical configurations $\boldsymbol{\sigma}$, and trained on the hidden representations $\boldsymbol{z}_p$, which are generated by a pretrained ViT at $J_2/J_1=0.4$.
As depicted in Fig.~\ref{fig:comp_MSR}, using the hidden representations (green curve) achieves an accuracy of $\Delta \varepsilon \approx 10^{-3}$, which is two orders of magnitude higher compared to the same network defined directly on the physical configurations ($\Delta \varepsilon \approx 10^{-1}$, orange curve). The difference primarily arises from the physical properties of the system that are encoded in the hidden representations, such as sign structure, amplitudes, and symmetries.
Furthermore, given that the sign structure at $J_2/J_1 = 0.4$ is well approximated by the MSR, we optimize an RBM, directly on the physical configurations, but implementing the Marshall sign prior (blue curve). This RBM achieves an accuracy of $\Delta \varepsilon \approx 10^{-2}$, underscoring that the information compressed in the hidden representation exceeds that provided by the Marshall sign prior. Despite increasing the number of parameters in RBMs, their performance remains inferior to the fine-tuned network due to the poor scaling behavior of the relative error in energy with the growth of network parameters and complicated structure of the landscape with a lot of local minima emerging when increasing the number hidden neurons (refer to the inset of Fig.~\ref{fig:comp_MSR}).
\bibliography{refs}

\begin{thebibliography}{56}%
\makeatletter
\providecommand \@ifxundefined [1]{%
 \@ifx{#1\undefined}
}%
\providecommand \@ifnum [1]{%
 \ifnum #1\expandafter \@firstoftwo
 \else \expandafter \@secondoftwo
 \fi
}%
\providecommand \@ifx [1]{%
 \ifx #1\expandafter \@firstoftwo
 \else \expandafter \@secondoftwo
 \fi
}%
\providecommand \natexlab [1]{#1}%
\providecommand \enquote  [1]{``#1''}%
\providecommand \bibnamefont  [1]{#1}%
\providecommand \bibfnamefont [1]{#1}%
\providecommand \citenamefont [1]{#1}%
\providecommand \href@noop [0]{\@secondoftwo}%
\providecommand \href [0]{\begingroup \@sanitize@url \@href}%
\providecommand \@href[1]{\@@startlink{#1}\@@href}%
\providecommand \@@href[1]{\endgroup#1\@@endlink}%
\providecommand \@sanitize@url [0]{\catcode `\\12\catcode `\$12\catcode `\&12\catcode `\#12\catcode `\^12\catcode `\_12\catcode `\%12\relax}%
\providecommand \@@startlink[1]{}%
\providecommand \@@endlink[0]{}%
\providecommand \url  [0]{\begingroup\@sanitize@url \@url }%
\providecommand \@url [1]{\endgroup\@href {#1}{\urlprefix }}%
\providecommand \urlprefix  [0]{URL }%
\providecommand \Eprint [0]{\href }%
\providecommand \doibase [0]{https://doi.org/}%
\providecommand \selectlanguage [0]{\@gobble}%
\providecommand \bibinfo  [0]{\@secondoftwo}%
\providecommand \bibfield  [0]{\@secondoftwo}%
\providecommand \translation [1]{[#1]}%
\providecommand \BibitemOpen [0]{}%
\providecommand \bibitemStop [0]{}%
\providecommand \bibitemNoStop [0]{.\EOS\space}%
\providecommand \EOS [0]{\spacefactor3000\relax}%
\providecommand \BibitemShut  [1]{\csname bibitem#1\endcsname}%
\let\auto@bib@innerbib\@empty
\bibitem [{\citenamefont {Bengio}\ \emph {et~al.}(2014)\citenamefont {Bengio}, \citenamefont {Courville},\ and\ \citenamefont {Vincent}}]{bengio2014}%
  \BibitemOpen
  \bibfield  {author} {\bibinfo {author} {\bibfnamefont {Y.}~\bibnamefont {Bengio}}, \bibinfo {author} {\bibfnamefont {A.}~\bibnamefont {Courville}},\ and\ \bibinfo {author} {\bibfnamefont {P.}~\bibnamefont {Vincent}},\ }\href@noop {} {\bibinfo {title} {Representation learning: A review and new perspectives}} (\bibinfo {year} {2014}),\ \Eprint {https://arxiv.org/abs/1206.5538} {arXiv:1206.5538 [cs.LG]} \BibitemShut {NoStop}%
\bibitem [{\citenamefont {LeCun}\ \emph {et~al.}(2015)\citenamefont {LeCun}, \citenamefont {Bengio},\ and\ \citenamefont {Hinton}}]{lecun2015deep}%
  \BibitemOpen
  \bibfield  {author} {\bibinfo {author} {\bibfnamefont {Y.}~\bibnamefont {LeCun}}, \bibinfo {author} {\bibfnamefont {Y.}~\bibnamefont {Bengio}},\ and\ \bibinfo {author} {\bibfnamefont {G.}~\bibnamefont {Hinton}},\ }\bibfield  {title} {\bibinfo {title} {Deep learning},\ }\href@noop {} {\bibfield  {journal} {\bibinfo  {journal} {Nature}\ }\textbf {\bibinfo {volume} {521}},\ \bibinfo {pages} {436} (\bibinfo {year} {2015})}\BibitemShut {NoStop}%
\bibitem [{\citenamefont {Vaswani}\ \emph {et~al.}(2017)\citenamefont {Vaswani}, \citenamefont {Shazeer}, \citenamefont {Parmar}, \citenamefont {Uszkoreit}, \citenamefont {Jones}, \citenamefont {Gomez}, \citenamefont {Kaiser},\ and\ \citenamefont {Polosukhin}}]{vaswani2017}%
  \BibitemOpen
  \bibfield  {author} {\bibinfo {author} {\bibfnamefont {A.}~\bibnamefont {Vaswani}}, \bibinfo {author} {\bibfnamefont {N.}~\bibnamefont {Shazeer}}, \bibinfo {author} {\bibfnamefont {N.}~\bibnamefont {Parmar}}, \bibinfo {author} {\bibfnamefont {J.}~\bibnamefont {Uszkoreit}}, \bibinfo {author} {\bibfnamefont {L.}~\bibnamefont {Jones}}, \bibinfo {author} {\bibfnamefont {A.}~\bibnamefont {Gomez}}, \bibinfo {author} {\bibfnamefont {L.}~\bibnamefont {Kaiser}},\ and\ \bibinfo {author} {\bibfnamefont {I.}~\bibnamefont {Polosukhin}},\ }\href {https://doi.org/10.48550/arXiv.1706.03762} {\bibinfo {title} {Attention is all you need}} (\bibinfo {year} {2017})\BibitemShut {NoStop}%
\bibitem [{\citenamefont {Radford}\ \emph {et~al.}(2019)\citenamefont {Radford}, \citenamefont {Wu}, \citenamefont {Child}, \citenamefont {Luan}, \citenamefont {Amodei}, \citenamefont {Sutskever} \emph {et~al.}}]{radford2019language}%
  \BibitemOpen
  \bibfield  {author} {\bibinfo {author} {\bibfnamefont {A.}~\bibnamefont {Radford}}, \bibinfo {author} {\bibfnamefont {J.}~\bibnamefont {Wu}}, \bibinfo {author} {\bibfnamefont {R.}~\bibnamefont {Child}}, \bibinfo {author} {\bibfnamefont {D.}~\bibnamefont {Luan}}, \bibinfo {author} {\bibfnamefont {D.}~\bibnamefont {Amodei}}, \bibinfo {author} {\bibfnamefont {I.}~\bibnamefont {Sutskever}}, \emph {et~al.},\ }\bibfield  {title} {\bibinfo {title} {Language models are unsupervised multitask learners},\ }\href@noop {} {\bibfield  {journal} {\bibinfo  {journal} {OpenAI blog}\ }\textbf {\bibinfo {volume} {1}},\ \bibinfo {pages} {9} (\bibinfo {year} {2019})}\BibitemShut {NoStop}%
\bibitem [{\citenamefont {Brown}\ \emph {et~al.}(2020)\citenamefont {Brown}, \citenamefont {Mann}, \citenamefont {Ryder}, \citenamefont {Subbiah}, \citenamefont {Kaplan}, \citenamefont {Dhariwal}, \citenamefont {Neelakantan}, \citenamefont {Shyam}, \citenamefont {Sastry}, \citenamefont {Askell}, \citenamefont {Agarwal}, \citenamefont {Herbert-Voss}, \citenamefont {Krueger}, \citenamefont {Henighan}, \citenamefont {Child}, \citenamefont {Ramesh}, \citenamefont {Ziegler}, \citenamefont {Wu}, \citenamefont {Winter}, \citenamefont {Hesse}, \citenamefont {Chen}, \citenamefont {Sigler}, \citenamefont {Litwin}, \citenamefont {Gray}, \citenamefont {Chess}, \citenamefont {Clark}, \citenamefont {Berner}, \citenamefont {McCandlish}, \citenamefont {Radford}, \citenamefont {Sutskever},\ and\ \citenamefont {Amodei}}]{gpt3}%
  \BibitemOpen
  \bibfield  {author} {\bibinfo {author} {\bibfnamefont {T.}~\bibnamefont {Brown}}, \bibinfo {author} {\bibfnamefont {B.}~\bibnamefont {Mann}}, \bibinfo {author} {\bibfnamefont {N.}~\bibnamefont {Ryder}}, \bibinfo {author} {\bibfnamefont {M.}~\bibnamefont {Subbiah}}, \bibinfo {author} {\bibfnamefont {J.~D.}\ \bibnamefont {Kaplan}}, \bibinfo {author} {\bibfnamefont {P.}~\bibnamefont {Dhariwal}}, \bibinfo {author} {\bibfnamefont {A.}~\bibnamefont {Neelakantan}}, \bibinfo {author} {\bibfnamefont {P.}~\bibnamefont {Shyam}}, \bibinfo {author} {\bibfnamefont {G.}~\bibnamefont {Sastry}}, \bibinfo {author} {\bibfnamefont {A.}~\bibnamefont {Askell}}, \bibinfo {author} {\bibfnamefont {S.}~\bibnamefont {Agarwal}}, \bibinfo {author} {\bibfnamefont {A.}~\bibnamefont {Herbert-Voss}}, \bibinfo {author} {\bibfnamefont {G.}~\bibnamefont {Krueger}}, \bibinfo {author} {\bibfnamefont {T.}~\bibnamefont {Henighan}}, \bibinfo {author} {\bibfnamefont {R.}~\bibnamefont {Child}}, \bibinfo {author} {\bibfnamefont {A.}~\bibnamefont
  {Ramesh}}, \bibinfo {author} {\bibfnamefont {D.}~\bibnamefont {Ziegler}}, \bibinfo {author} {\bibfnamefont {J.}~\bibnamefont {Wu}}, \bibinfo {author} {\bibfnamefont {C.}~\bibnamefont {Winter}}, \bibinfo {author} {\bibfnamefont {C.}~\bibnamefont {Hesse}}, \bibinfo {author} {\bibfnamefont {M.}~\bibnamefont {Chen}}, \bibinfo {author} {\bibfnamefont {E.}~\bibnamefont {Sigler}}, \bibinfo {author} {\bibfnamefont {M.}~\bibnamefont {Litwin}}, \bibinfo {author} {\bibfnamefont {S.}~\bibnamefont {Gray}}, \bibinfo {author} {\bibfnamefont {B.}~\bibnamefont {Chess}}, \bibinfo {author} {\bibfnamefont {J.}~\bibnamefont {Clark}}, \bibinfo {author} {\bibfnamefont {C.}~\bibnamefont {Berner}}, \bibinfo {author} {\bibfnamefont {S.}~\bibnamefont {McCandlish}}, \bibinfo {author} {\bibfnamefont {A.}~\bibnamefont {Radford}}, \bibinfo {author} {\bibfnamefont {I.}~\bibnamefont {Sutskever}},\ and\ \bibinfo {author} {\bibfnamefont {D.}~\bibnamefont {Amodei}},\ }\bibfield  {title} {\bibinfo {title} {Language models are few-shot
  learners},\ }in\ \href {https://proceedings.neurips.cc/paper_files/paper/2020/file/1457c0d6bfcb4967418bfb8ac142f64a-Paper.pdf} {\emph {\bibinfo {booktitle} {Advances in Neural Information Processing Systems}}},\ Vol.~\bibinfo {volume} {33},\ \bibinfo {editor} {edited by\ \bibinfo {editor} {\bibfnamefont {H.}~\bibnamefont {Larochelle}}, \bibinfo {editor} {\bibfnamefont {M.}~\bibnamefont {Ranzato}}, \bibinfo {editor} {\bibfnamefont {R.}~\bibnamefont {Hadsell}}, \bibinfo {editor} {\bibfnamefont {M.}~\bibnamefont {Balcan}},\ and\ \bibinfo {editor} {\bibfnamefont {H.}~\bibnamefont {Lin}}}\ (\bibinfo  {publisher} {Curran Associates, Inc.},\ \bibinfo {year} {2020})\ pp.\ \bibinfo {pages} {1877--1901}\BibitemShut {NoStop}%
\bibitem [{\citenamefont {Dosovitskiy}\ \emph {et~al.}(2021)\citenamefont {Dosovitskiy}, \citenamefont {Beyer}, \citenamefont {Kolesnikov}, \citenamefont {Weissenborn}, \citenamefont {Zhai}, \citenamefont {Unterthiner}, \citenamefont {Dehghani}, \citenamefont {Minderer}, \citenamefont {Heigold}, \citenamefont {Gelly}, \citenamefont {Uszkoreit},\ and\ \citenamefont {Houlsby}}]{dosovitskiy2021}%
  \BibitemOpen
  \bibfield  {author} {\bibinfo {author} {\bibfnamefont {A.}~\bibnamefont {Dosovitskiy}}, \bibinfo {author} {\bibfnamefont {L.}~\bibnamefont {Beyer}}, \bibinfo {author} {\bibfnamefont {A.}~\bibnamefont {Kolesnikov}}, \bibinfo {author} {\bibfnamefont {D.}~\bibnamefont {Weissenborn}}, \bibinfo {author} {\bibfnamefont {X.}~\bibnamefont {Zhai}}, \bibinfo {author} {\bibfnamefont {T.}~\bibnamefont {Unterthiner}}, \bibinfo {author} {\bibfnamefont {M.}~\bibnamefont {Dehghani}}, \bibinfo {author} {\bibfnamefont {M.}~\bibnamefont {Minderer}}, \bibinfo {author} {\bibfnamefont {G.}~\bibnamefont {Heigold}}, \bibinfo {author} {\bibfnamefont {S.}~\bibnamefont {Gelly}}, \bibinfo {author} {\bibfnamefont {J.}~\bibnamefont {Uszkoreit}},\ and\ \bibinfo {author} {\bibfnamefont {N.}~\bibnamefont {Houlsby}},\ }\href {https://doi.org/10.48550/arXiv.2010.11929} {\bibinfo {title} {An image is worth 16x16 words: Transformers for image recognition at scale}} (\bibinfo {year} {2021})\BibitemShut {NoStop}%
\bibitem [{\citenamefont {Jumper}\ \emph {et~al.}(2021)\citenamefont {Jumper}, \citenamefont {Evans}, \citenamefont {Pritzel}, \citenamefont {Green}, \citenamefont {Figurnov}, \citenamefont {Ronneberger}, \citenamefont {Tunyasuvunakool}, \citenamefont {Bates}, \citenamefont {{\v{Z}}{\'\i}dek}, \citenamefont {Potapenko} \emph {et~al.}}]{jumper2021highly}%
  \BibitemOpen
  \bibfield  {author} {\bibinfo {author} {\bibfnamefont {J.}~\bibnamefont {Jumper}}, \bibinfo {author} {\bibfnamefont {R.}~\bibnamefont {Evans}}, \bibinfo {author} {\bibfnamefont {A.}~\bibnamefont {Pritzel}}, \bibinfo {author} {\bibfnamefont {T.}~\bibnamefont {Green}}, \bibinfo {author} {\bibfnamefont {M.}~\bibnamefont {Figurnov}}, \bibinfo {author} {\bibfnamefont {O.}~\bibnamefont {Ronneberger}}, \bibinfo {author} {\bibfnamefont {K.}~\bibnamefont {Tunyasuvunakool}}, \bibinfo {author} {\bibfnamefont {R.}~\bibnamefont {Bates}}, \bibinfo {author} {\bibfnamefont {A.}~\bibnamefont {{\v{Z}}{\'\i}dek}}, \bibinfo {author} {\bibfnamefont {A.}~\bibnamefont {Potapenko}}, \emph {et~al.},\ }\bibfield  {title} {\bibinfo {title} {Highly accurate protein structure prediction with alphafold},\ }\href {https://doi.org/https://doi.org/10.1038/s41586-021-03819-2} {\bibfield  {journal} {\bibinfo  {journal} {Nature}\ }\textbf {\bibinfo {volume} {596}},\ \bibinfo {pages} {583} (\bibinfo {year} {2021})}\BibitemShut {NoStop}%
\bibitem [{\citenamefont {Carleo}\ and\ \citenamefont {Troyer}(2017)}]{carleo2017}%
  \BibitemOpen
  \bibfield  {author} {\bibinfo {author} {\bibfnamefont {G.}~\bibnamefont {Carleo}}\ and\ \bibinfo {author} {\bibfnamefont {M.}~\bibnamefont {Troyer}},\ }\bibfield  {title} {\bibinfo {title} {Solving the quantum many-body problem with artificial neural networks},\ }\href {https://doi.org/10.1126/science.aag2302} {\bibfield  {journal} {\bibinfo  {journal} {Science}\ }\textbf {\bibinfo {volume} {355}},\ \bibinfo {pages} {602} (\bibinfo {year} {2017})}\BibitemShut {NoStop}%
\bibitem [{\citenamefont {Luo}\ and\ \citenamefont {Clark}(2019)}]{luo2019}%
  \BibitemOpen
  \bibfield  {author} {\bibinfo {author} {\bibfnamefont {D.}~\bibnamefont {Luo}}\ and\ \bibinfo {author} {\bibfnamefont {B.~K.}\ \bibnamefont {Clark}},\ }\bibfield  {title} {\bibinfo {title} {Backflow transformations via neural networks for quantum many-body wave functions},\ }\bibfield  {journal} {\bibinfo  {journal} {Physical Review Letters}\ }\textbf {\bibinfo {volume} {122}},\ \href {https://doi.org/10.1103/physrevlett.122.226401} {10.1103/physrevlett.122.226401} (\bibinfo {year} {2019})\BibitemShut {NoStop}%
\bibitem [{\citenamefont {Nomura}\ and\ \citenamefont {Imada}(2021)}]{nomuraimada2021}%
  \BibitemOpen
  \bibfield  {author} {\bibinfo {author} {\bibfnamefont {Y.}~\bibnamefont {Nomura}}\ and\ \bibinfo {author} {\bibfnamefont {M.}~\bibnamefont {Imada}},\ }\bibfield  {title} {\bibinfo {title} {Dirac-type nodal spin liquid revealed by refined quantum many-body solver using neural-network wave function, correlation ratio, and level spectroscopy},\ }\href {https://doi.org/10.1103/PhysRevX.11.031034} {\bibfield  {journal} {\bibinfo  {journal} {Phys. Rev. X}\ }\textbf {\bibinfo {volume} {11}},\ \bibinfo {pages} {031034} (\bibinfo {year} {2021})}\BibitemShut {NoStop}%
\bibitem [{\citenamefont {Robledo~Moreno}\ \emph {et~al.}(2022{\natexlab{a}})\citenamefont {Robledo~Moreno}, \citenamefont {Carleo}, \citenamefont {Georges},\ and\ \citenamefont {Stokes}}]{robledomoreno2022}%
  \BibitemOpen
  \bibfield  {author} {\bibinfo {author} {\bibfnamefont {J.}~\bibnamefont {Robledo~Moreno}}, \bibinfo {author} {\bibfnamefont {G.}~\bibnamefont {Carleo}}, \bibinfo {author} {\bibfnamefont {A.}~\bibnamefont {Georges}},\ and\ \bibinfo {author} {\bibfnamefont {J.}~\bibnamefont {Stokes}},\ }\bibfield  {title} {\bibinfo {title} {Fermionic wave functions from neural-network constrained hidden states},\ }\bibfield  {journal} {\bibinfo  {journal} {Proceedings of the National Academy of Sciences}\ }\textbf {\bibinfo {volume} {119}},\ \href {https://doi.org/10.1073/pnas.2122059119} {10.1073/pnas.2122059119} (\bibinfo {year} {2022}{\natexlab{a}})\BibitemShut {NoStop}%
\bibitem [{\citenamefont {Roth}\ \emph {et~al.}(2023)\citenamefont {Roth}, \citenamefont {Szab\'o},\ and\ \citenamefont {MacDonald}}]{roth2023}%
  \BibitemOpen
  \bibfield  {author} {\bibinfo {author} {\bibfnamefont {C.}~\bibnamefont {Roth}}, \bibinfo {author} {\bibfnamefont {A.}~\bibnamefont {Szab\'o}},\ and\ \bibinfo {author} {\bibfnamefont {A.~H.}\ \bibnamefont {MacDonald}},\ }\bibfield  {title} {\bibinfo {title} {High-accuracy variational monte carlo for frustrated magnets with deep neural networks},\ }\href {https://doi.org/10.1103/PhysRevB.108.054410} {\bibfield  {journal} {\bibinfo  {journal} {Phys. Rev. B}\ }\textbf {\bibinfo {volume} {108}},\ \bibinfo {pages} {054410} (\bibinfo {year} {2023})}\BibitemShut {NoStop}%
\bibitem [{\citenamefont {Lange}\ \emph {et~al.}(2024)\citenamefont {Lange}, \citenamefont {de~Walle}, \citenamefont {Abedinnia},\ and\ \citenamefont {Bohrdt}}]{lange2024architectures}%
  \BibitemOpen
  \bibfield  {author} {\bibinfo {author} {\bibfnamefont {H.}~\bibnamefont {Lange}}, \bibinfo {author} {\bibfnamefont {A.~V.}\ \bibnamefont {de~Walle}}, \bibinfo {author} {\bibfnamefont {A.}~\bibnamefont {Abedinnia}},\ and\ \bibinfo {author} {\bibfnamefont {A.}~\bibnamefont {Bohrdt}},\ }\href@noop {} {\bibinfo {title} {From architectures to applications: A review of neural quantum states}} (\bibinfo {year} {2024}),\ \Eprint {https://arxiv.org/abs/2402.09402} {arXiv:2402.09402 [cond-mat.dis-nn]} \BibitemShut {NoStop}%
\bibitem [{\citenamefont {Lange}\ \emph {et~al.}(2023)\citenamefont {Lange}, \citenamefont {D{\"o}schl}, \citenamefont {Carrasquilla},\ and\ \citenamefont {Bohrdt}}]{lange2023neural}%
  \BibitemOpen
  \bibfield  {author} {\bibinfo {author} {\bibfnamefont {H.}~\bibnamefont {Lange}}, \bibinfo {author} {\bibfnamefont {F.}~\bibnamefont {D{\"o}schl}}, \bibinfo {author} {\bibfnamefont {J.~F.}\ \bibnamefont {Carrasquilla}},\ and\ \bibinfo {author} {\bibfnamefont {A.}~\bibnamefont {Bohrdt}},\ }\bibfield  {title} {\bibinfo {title} {Neural network approach to quasiparticle dispersions in doped antiferromagnets},\ }\href@noop {} {\bibfield  {journal} {\bibinfo  {journal} {Communications Physics}\ }\textbf {\bibinfo {volume} {7}},\ \bibinfo {pages} {1} (\bibinfo {year} {2023})}\BibitemShut {NoStop}%
\bibitem [{\citenamefont {Pfau}\ \emph {et~al.}(2020)\citenamefont {Pfau}, \citenamefont {Spencer}, \citenamefont {Matthews},\ and\ \citenamefont {Foulkes}}]{pfau2020}%
  \BibitemOpen
  \bibfield  {author} {\bibinfo {author} {\bibfnamefont {D.}~\bibnamefont {Pfau}}, \bibinfo {author} {\bibfnamefont {J.~S.}\ \bibnamefont {Spencer}}, \bibinfo {author} {\bibfnamefont {A.~G. D.~G.}\ \bibnamefont {Matthews}},\ and\ \bibinfo {author} {\bibfnamefont {W.~M.~C.}\ \bibnamefont {Foulkes}},\ }\bibfield  {title} {\bibinfo {title} {Ab initio solution of the many-electron schr\"odinger equation with deep neural networks},\ }\href {https://doi.org/10.1103/PhysRevResearch.2.033429} {\bibfield  {journal} {\bibinfo  {journal} {Phys. Rev. Res.}\ }\textbf {\bibinfo {volume} {2}},\ \bibinfo {pages} {033429} (\bibinfo {year} {2020})}\BibitemShut {NoStop}%
\bibitem [{\citenamefont {Kim}\ \emph {et~al.}(2023)\citenamefont {Kim}, \citenamefont {Pescia}, \citenamefont {Fore}, \citenamefont {Nys}, \citenamefont {Carleo}, \citenamefont {Gandolfi}, \citenamefont {Hjorth-Jensen},\ and\ \citenamefont {Lovato}}]{kim2023}%
  \BibitemOpen
  \bibfield  {author} {\bibinfo {author} {\bibfnamefont {J.}~\bibnamefont {Kim}}, \bibinfo {author} {\bibfnamefont {G.}~\bibnamefont {Pescia}}, \bibinfo {author} {\bibfnamefont {B.}~\bibnamefont {Fore}}, \bibinfo {author} {\bibfnamefont {J.}~\bibnamefont {Nys}}, \bibinfo {author} {\bibfnamefont {G.}~\bibnamefont {Carleo}}, \bibinfo {author} {\bibfnamefont {S.}~\bibnamefont {Gandolfi}}, \bibinfo {author} {\bibfnamefont {M.}~\bibnamefont {Hjorth-Jensen}},\ and\ \bibinfo {author} {\bibfnamefont {A.}~\bibnamefont {Lovato}},\ }\href@noop {} {\bibinfo {title} {Neural-network quantum states for ultra-cold fermi gases}} (\bibinfo {year} {2023}),\ \Eprint {https://arxiv.org/abs/2305.08831} {arXiv:2305.08831 [cond-mat.quant-gas]} \BibitemShut {NoStop}%
\bibitem [{\citenamefont {Pescia}\ \emph {et~al.}(2022)\citenamefont {Pescia}, \citenamefont {Han}, \citenamefont {Lovato}, \citenamefont {Lu},\ and\ \citenamefont {Carleo}}]{pescia2022}%
  \BibitemOpen
  \bibfield  {author} {\bibinfo {author} {\bibfnamefont {G.}~\bibnamefont {Pescia}}, \bibinfo {author} {\bibfnamefont {J.}~\bibnamefont {Han}}, \bibinfo {author} {\bibfnamefont {A.}~\bibnamefont {Lovato}}, \bibinfo {author} {\bibfnamefont {J.}~\bibnamefont {Lu}},\ and\ \bibinfo {author} {\bibfnamefont {G.}~\bibnamefont {Carleo}},\ }\bibfield  {title} {\bibinfo {title} {Neural-network quantum states for periodic systems in continuous space},\ }\href {https://doi.org/10.1103/PhysRevResearch.4.023138} {\bibfield  {journal} {\bibinfo  {journal} {Phys. Rev. Res.}\ }\textbf {\bibinfo {volume} {4}},\ \bibinfo {pages} {023138} (\bibinfo {year} {2022})}\BibitemShut {NoStop}%
\bibitem [{\citenamefont {Robledo~Moreno}\ \emph {et~al.}(2022{\natexlab{b}})\citenamefont {Robledo~Moreno}, \citenamefont {Carleo}, \citenamefont {Georges},\ and\ \citenamefont {Stokes}}]{moreno2022}%
  \BibitemOpen
  \bibfield  {author} {\bibinfo {author} {\bibfnamefont {J.}~\bibnamefont {Robledo~Moreno}}, \bibinfo {author} {\bibfnamefont {G.}~\bibnamefont {Carleo}}, \bibinfo {author} {\bibfnamefont {A.}~\bibnamefont {Georges}},\ and\ \bibinfo {author} {\bibfnamefont {J.}~\bibnamefont {Stokes}},\ }\bibfield  {title} {\bibinfo {title} {Fermionic wave functions from neural-network constrained hidden states},\ }\bibfield  {journal} {\bibinfo  {journal} {Proceedings of the National Academy of Sciences}\ }\textbf {\bibinfo {volume} {119}},\ \href {https://doi.org/10.1073/pnas.2122059119} {10.1073/pnas.2122059119} (\bibinfo {year} {2022}{\natexlab{b}})\BibitemShut {NoStop}%
\bibitem [{\citenamefont {Becca}\ and\ \citenamefont {Sorella}(2017)}]{becca2017}%
  \BibitemOpen
  \bibfield  {author} {\bibinfo {author} {\bibfnamefont {F.}~\bibnamefont {Becca}}\ and\ \bibinfo {author} {\bibfnamefont {S.}~\bibnamefont {Sorella}},\ }\href {https://doi.org/10.1017/9781316417041} {\emph {\bibinfo {title} {Quantum Monte Carlo Approaches for Correlated Systems}}}\ (\bibinfo  {publisher} {Cambridge University Press},\ \bibinfo {year} {2017})\BibitemShut {NoStop}%
\bibitem [{\citenamefont {Viteritti}\ \emph {et~al.}(2023{\natexlab{a}})\citenamefont {Viteritti}, \citenamefont {Rende}, \citenamefont {Parola}, \citenamefont {Goldt},\ and\ \citenamefont {Becca}}]{viteritti2023}%
  \BibitemOpen
  \bibfield  {author} {\bibinfo {author} {\bibfnamefont {L.~L.}\ \bibnamefont {Viteritti}}, \bibinfo {author} {\bibfnamefont {R.}~\bibnamefont {Rende}}, \bibinfo {author} {\bibfnamefont {A.}~\bibnamefont {Parola}}, \bibinfo {author} {\bibfnamefont {S.}~\bibnamefont {Goldt}},\ and\ \bibinfo {author} {\bibfnamefont {F.}~\bibnamefont {Becca}},\ }\href@noop {} {\bibinfo {title} {Transformer wave function for the shastry-sutherland model: emergence of a spin-liquid phase}} (\bibinfo {year} {2023}{\natexlab{a}}),\ \Eprint {https://arxiv.org/abs/2311.16889} {arXiv:2311.16889 [cond-mat.str-el]} \BibitemShut {NoStop}%
\bibitem [{\citenamefont {Rende}\ \emph {et~al.}(2024{\natexlab{a}})\citenamefont {Rende}, \citenamefont {Viteritti}, \citenamefont {Bardone}, \citenamefont {Becca},\ and\ \citenamefont {Goldt}}]{rende2023}%
  \BibitemOpen
  \bibfield  {author} {\bibinfo {author} {\bibfnamefont {R.}~\bibnamefont {Rende}}, \bibinfo {author} {\bibfnamefont {L.~L.}\ \bibnamefont {Viteritti}}, \bibinfo {author} {\bibfnamefont {L.}~\bibnamefont {Bardone}}, \bibinfo {author} {\bibfnamefont {F.}~\bibnamefont {Becca}},\ and\ \bibinfo {author} {\bibfnamefont {S.}~\bibnamefont {Goldt}},\ }\bibfield  {title} {\bibinfo {title} {A simple linear algebra identity to optimize large-scale neural network quantum states},\ }\bibfield  {journal} {\bibinfo  {journal} {Communications Physics}\ }\textbf {\bibinfo {volume} {7}},\ \href {https://doi.org/10.1038/s42005-024-01732-4} {10.1038/s42005-024-01732-4} (\bibinfo {year} {2024}{\natexlab{a}})\BibitemShut {NoStop}%
\bibitem [{\citenamefont {Tang}\ \emph {et~al.}(2024)\citenamefont {Tang}, \citenamefont {Liu}, \citenamefont {Zhang},\ and\ \citenamefont {Zhang}}]{tang2024learning}%
  \BibitemOpen
  \bibfield  {author} {\bibinfo {author} {\bibfnamefont {Y.}~\bibnamefont {Tang}}, \bibinfo {author} {\bibfnamefont {J.}~\bibnamefont {Liu}}, \bibinfo {author} {\bibfnamefont {J.}~\bibnamefont {Zhang}},\ and\ \bibinfo {author} {\bibfnamefont {P.}~\bibnamefont {Zhang}},\ }\bibfield  {title} {\bibinfo {title} {Learning nonequilibrium statistical mechanics and dynamical phase transitions},\ }\href {https://doi.org/https://doi.org/10.1038/s41467-024-45172-8} {\bibfield  {journal} {\bibinfo  {journal} {Nature Communications}\ }\textbf {\bibinfo {volume} {15}},\ \bibinfo {pages} {1117} (\bibinfo {year} {2024})}\BibitemShut {NoStop}%
\bibitem [{\citenamefont {Melko}\ and\ \citenamefont {Carrasquilla}(2024)}]{melko2024language}%
  \BibitemOpen
  \bibfield  {author} {\bibinfo {author} {\bibfnamefont {R.~G.}\ \bibnamefont {Melko}}\ and\ \bibinfo {author} {\bibfnamefont {J.}~\bibnamefont {Carrasquilla}},\ }\bibfield  {title} {\bibinfo {title} {{Language models for quantum simulation}},\ }\href {https://doi.org/10.1038/s43588-023-00578-0} {\bibfield  {journal} {\bibinfo  {journal} {Nature Computat. Sci.}\ }\textbf {\bibinfo {volume} {4}},\ \bibinfo {pages} {11} (\bibinfo {year} {2024})}\BibitemShut {NoStop}%
\bibitem [{\citenamefont {Sprague}\ and\ \citenamefont {Czischek}(2024)}]{czischek2023}%
  \BibitemOpen
  \bibfield  {author} {\bibinfo {author} {\bibfnamefont {K.}~\bibnamefont {Sprague}}\ and\ \bibinfo {author} {\bibfnamefont {S.}~\bibnamefont {Czischek}},\ }\bibfield  {title} {\bibinfo {title} {Variational monte carlo with large patched transformers},\ }\bibfield  {journal} {\bibinfo  {journal} {Communications Physics}\ }\textbf {\bibinfo {volume} {7}},\ \href {https://doi.org/10.1038/s42005-024-01584-y} {10.1038/s42005-024-01584-y} (\bibinfo {year} {2024})\BibitemShut {NoStop}%
\bibitem [{\citenamefont {Luo}\ \emph {et~al.}(2022)\citenamefont {Luo}, \citenamefont {Chen}, \citenamefont {Carrasquilla},\ and\ \citenamefont {Clark}}]{luo2022}%
  \BibitemOpen
  \bibfield  {author} {\bibinfo {author} {\bibfnamefont {D.}~\bibnamefont {Luo}}, \bibinfo {author} {\bibfnamefont {Z.}~\bibnamefont {Chen}}, \bibinfo {author} {\bibfnamefont {J.}~\bibnamefont {Carrasquilla}},\ and\ \bibinfo {author} {\bibfnamefont {B.~K.}\ \bibnamefont {Clark}},\ }\bibfield  {title} {\bibinfo {title} {Autoregressive neural network for simulating open quantum systems via a probabilistic formulation},\ }\href {https://doi.org/10.1103/PhysRevLett.128.090501} {\bibfield  {journal} {\bibinfo  {journal} {Phys. Rev. Lett.}\ }\textbf {\bibinfo {volume} {128}},\ \bibinfo {pages} {090501} (\bibinfo {year} {2022})}\BibitemShut {NoStop}%
\bibitem [{\citenamefont {Luo}\ \emph {et~al.}(2023)\citenamefont {Luo}, \citenamefont {Chen}, \citenamefont {Hu}, \citenamefont {Zhao}, \citenamefont {Hur},\ and\ \citenamefont {Clark}}]{diluo2023}%
  \BibitemOpen
  \bibfield  {author} {\bibinfo {author} {\bibfnamefont {D.}~\bibnamefont {Luo}}, \bibinfo {author} {\bibfnamefont {Z.}~\bibnamefont {Chen}}, \bibinfo {author} {\bibfnamefont {K.}~\bibnamefont {Hu}}, \bibinfo {author} {\bibfnamefont {Z.}~\bibnamefont {Zhao}}, \bibinfo {author} {\bibfnamefont {V.~M.}\ \bibnamefont {Hur}},\ and\ \bibinfo {author} {\bibfnamefont {B.~K.}\ \bibnamefont {Clark}},\ }\bibfield  {title} {\bibinfo {title} {Gauge-invariant and anyonic-symmetric autoregressive neural network for quantum lattice models},\ }\href {https://doi.org/10.1103/PhysRevResearch.5.013216} {\bibfield  {journal} {\bibinfo  {journal} {Phys. Rev. Res.}\ }\textbf {\bibinfo {volume} {5}},\ \bibinfo {pages} {013216} (\bibinfo {year} {2023})}\BibitemShut {NoStop}%
\bibitem [{\citenamefont {Viteritti}\ \emph {et~al.}(2023{\natexlab{b}})\citenamefont {Viteritti}, \citenamefont {Rende},\ and\ \citenamefont {Becca}}]{viteritti20231d}%
  \BibitemOpen
  \bibfield  {author} {\bibinfo {author} {\bibfnamefont {L.~L.}\ \bibnamefont {Viteritti}}, \bibinfo {author} {\bibfnamefont {R.}~\bibnamefont {Rende}},\ and\ \bibinfo {author} {\bibfnamefont {F.}~\bibnamefont {Becca}},\ }\bibfield  {title} {\bibinfo {title} {Transformer variational wave functions for frustrated quantum spin systems},\ }\href {https://doi.org/10.1103/PhysRevLett.130.236401} {\bibfield  {journal} {\bibinfo  {journal} {Phys. Rev. Lett.}\ }\textbf {\bibinfo {volume} {130}},\ \bibinfo {pages} {236401} (\bibinfo {year} {2023}{\natexlab{b}})}\BibitemShut {NoStop}%
\bibitem [{\citenamefont {Rende}\ \emph {et~al.}(2024{\natexlab{b}})\citenamefont {Rende}, \citenamefont {Gerace}, \citenamefont {Laio},\ and\ \citenamefont {Goldt}}]{rende2023b}%
  \BibitemOpen
  \bibfield  {author} {\bibinfo {author} {\bibfnamefont {R.}~\bibnamefont {Rende}}, \bibinfo {author} {\bibfnamefont {F.}~\bibnamefont {Gerace}}, \bibinfo {author} {\bibfnamefont {A.}~\bibnamefont {Laio}},\ and\ \bibinfo {author} {\bibfnamefont {S.}~\bibnamefont {Goldt}},\ }\bibfield  {title} {\bibinfo {title} {Mapping of attention mechanisms to a generalized potts model},\ }\href {https://doi.org/10.1103/PhysRevResearch.6.023057} {\bibfield  {journal} {\bibinfo  {journal} {Phys. Rev. Res.}\ }\textbf {\bibinfo {volume} {6}},\ \bibinfo {pages} {023057} (\bibinfo {year} {2024}{\natexlab{b}})}\BibitemShut {NoStop}%
\bibitem [{\citenamefont {Sorella}(2005)}]{sorella2005}%
  \BibitemOpen
  \bibfield  {author} {\bibinfo {author} {\bibfnamefont {S.}~\bibnamefont {Sorella}},\ }\bibfield  {title} {\bibinfo {title} {Wave function optimization in the variational monte carlo method},\ }\href {https://doi.org/10.1103/PhysRevB.71.241103} {\bibfield  {journal} {\bibinfo  {journal} {Phys. Rev. B}\ }\textbf {\bibinfo {volume} {71}},\ \bibinfo {pages} {241103} (\bibinfo {year} {2005})}\BibitemShut {NoStop}%
\bibitem [{\citenamefont {Chen}\ and\ \citenamefont {Heyl}(2023)}]{chen2023}%
  \BibitemOpen
  \bibfield  {author} {\bibinfo {author} {\bibfnamefont {A.}~\bibnamefont {Chen}}\ and\ \bibinfo {author} {\bibfnamefont {M.}~\bibnamefont {Heyl}},\ }\href@noop {} {\bibinfo {title} {Efficient optimization of deep neural quantum states toward machine precision}} (\bibinfo {year} {2023}),\ \Eprint {https://arxiv.org/abs/2302.01941} {arXiv:2302.01941 [cond-mat.dis-nn]} \BibitemShut {NoStop}%
\bibitem [{\citenamefont {Vicentini}\ \emph {et~al.}(2022)\citenamefont {Vicentini}, \citenamefont {Hofmann}, \citenamefont {Szabó}, \citenamefont {Wu}, \citenamefont {Roth}, \citenamefont {Giuliani}, \citenamefont {Pescia}, \citenamefont {Nys}, \citenamefont {Vargas-Calderón}, \citenamefont {Astrakhantsev},\ and\ \citenamefont {Carleo}}]{netket3}%
  \BibitemOpen
  \bibfield  {author} {\bibinfo {author} {\bibfnamefont {F.}~\bibnamefont {Vicentini}}, \bibinfo {author} {\bibfnamefont {D.}~\bibnamefont {Hofmann}}, \bibinfo {author} {\bibfnamefont {A.}~\bibnamefont {Szabó}}, \bibinfo {author} {\bibfnamefont {D.}~\bibnamefont {Wu}}, \bibinfo {author} {\bibfnamefont {C.}~\bibnamefont {Roth}}, \bibinfo {author} {\bibfnamefont {C.}~\bibnamefont {Giuliani}}, \bibinfo {author} {\bibfnamefont {G.}~\bibnamefont {Pescia}}, \bibinfo {author} {\bibfnamefont {J.}~\bibnamefont {Nys}}, \bibinfo {author} {\bibfnamefont {V.}~\bibnamefont {Vargas-Calderón}}, \bibinfo {author} {\bibfnamefont {N.}~\bibnamefont {Astrakhantsev}},\ and\ \bibinfo {author} {\bibfnamefont {G.}~\bibnamefont {Carleo}},\ }\bibfield  {title} {\bibinfo {title} {{NetKet 3: Machine Learning Toolbox for Many-Body Quantum Systems}},\ }\href {https://doi.org/10.21468/SciPostPhysCodeb.7} {\bibfield  {journal} {\bibinfo  {journal} {SciPost Phys. Codebases}\ ,\ \bibinfo {pages} {7}} (\bibinfo {year} {2022})}\BibitemShut
  {NoStop}%
\bibitem [{\citenamefont {White}(1992)}]{white1992}%
  \BibitemOpen
  \bibfield  {author} {\bibinfo {author} {\bibfnamefont {S.~R.}\ \bibnamefont {White}},\ }\bibfield  {title} {\bibinfo {title} {Density matrix formulation for quantum renormalization groups},\ }\href {https://doi.org/10.1103/PhysRevLett.69.2863} {\bibfield  {journal} {\bibinfo  {journal} {Phys. Rev. Lett.}\ }\textbf {\bibinfo {volume} {69}},\ \bibinfo {pages} {2863} (\bibinfo {year} {1992})}\BibitemShut {NoStop}%
\bibitem [{Note1()}]{Note1}%
  \BibitemOpen
  \bibinfo {note} {UMAP~\cite {mcinnes2020} is a general purpose dimension reduction technique for machine learning. It is constructed from a theoretical framework based in Riemannian geometry and algebraic topology, see Ref.~\cite {mcinnes2020} for more details.}\BibitemShut {Stop}%
\bibitem [{\citenamefont {Eggert}(1996)}]{eggert1996}%
  \BibitemOpen
  \bibfield  {author} {\bibinfo {author} {\bibfnamefont {S.}~\bibnamefont {Eggert}},\ }\bibfield  {title} {\bibinfo {title} {Numerical evidence for multiplicative logarithmic corrections from marginal operators},\ }\href {https://doi.org/10.1103/PhysRevB.54.R9612} {\bibfield  {journal} {\bibinfo  {journal} {Phys. Rev. B}\ }\textbf {\bibinfo {volume} {54}},\ \bibinfo {pages} {R9612} (\bibinfo {year} {1996})}\BibitemShut {NoStop}%
\bibitem [{\citenamefont {Lacroix}\ \emph {et~al.}(2011)\citenamefont {Lacroix}, \citenamefont {Mendels},\ and\ \citenamefont {Mila}}]{lacroix2011book}%
  \BibitemOpen
  \bibfield  {author} {\bibinfo {author} {\bibfnamefont {C.}~\bibnamefont {Lacroix}}, \bibinfo {author} {\bibfnamefont {P.}~\bibnamefont {Mendels}},\ and\ \bibinfo {author} {\bibfnamefont {F.}~\bibnamefont {Mila}},\ }\href {https://doi.org/10.1007/978-3-642-10589-0} {\emph {\bibinfo {title} {Introduction to Frustrated Magnetism: Materials, Experiments, Theory}}}\ (\bibinfo {year} {2011})\BibitemShut {NoStop}%
\bibitem [{\citenamefont {Capriotti}\ \emph {et~al.}(2003)\citenamefont {Capriotti}, \citenamefont {Becca}, \citenamefont {Parola},\ and\ \citenamefont {Sorella}}]{capriotti2003}%
  \BibitemOpen
  \bibfield  {author} {\bibinfo {author} {\bibfnamefont {L.}~\bibnamefont {Capriotti}}, \bibinfo {author} {\bibfnamefont {F.}~\bibnamefont {Becca}}, \bibinfo {author} {\bibfnamefont {A.}~\bibnamefont {Parola}},\ and\ \bibinfo {author} {\bibfnamefont {S.}~\bibnamefont {Sorella}},\ }\bibfield  {title} {\bibinfo {title} {Suppression of dimer correlations in the two-dimensional ${J}_{1}\ensuremath{-}{J}_{2}$ heisenberg model: An exact diagonalization study},\ }\href {https://doi.org/10.1103/PhysRevB.67.212402} {\bibfield  {journal} {\bibinfo  {journal} {Phys. Rev. B}\ }\textbf {\bibinfo {volume} {67}},\ \bibinfo {pages} {212402} (\bibinfo {year} {2003})}\BibitemShut {NoStop}%
\bibitem [{\citenamefont {Marshall}(1955)}]{marshall1955}%
  \BibitemOpen
  \bibfield  {author} {\bibinfo {author} {\bibfnamefont {W.}~\bibnamefont {Marshall}},\ }\bibfield  {title} {\bibinfo {title} {Antiferromagnetism},\ }\href {http://www.jstor.org/stable/99682} {\bibfield  {journal} {\bibinfo  {journal} {Proceedings of the Royal Society of London. Series A, Mathematical and Physical Sciences}\ }\textbf {\bibinfo {volume} {232}},\ \bibinfo {pages} {48} (\bibinfo {year} {1955})}\BibitemShut {NoStop}%
\bibitem [{\citenamefont {Viteritti}\ \emph {et~al.}(2022)\citenamefont {Viteritti}, \citenamefont {Ferrari},\ and\ \citenamefont {Becca}}]{viteritti2022}%
  \BibitemOpen
  \bibfield  {author} {\bibinfo {author} {\bibfnamefont {L.~L.}\ \bibnamefont {Viteritti}}, \bibinfo {author} {\bibfnamefont {F.}~\bibnamefont {Ferrari}},\ and\ \bibinfo {author} {\bibfnamefont {F.}~\bibnamefont {Becca}},\ }\bibfield  {title} {\bibinfo {title} {{Accuracy of restricted Boltzmann machines for the one-dimensional $J_1-J_2$ Heisenberg model}},\ }\href {https://doi.org/10.21468/SciPostPhys.12.5.166} {\bibfield  {journal} {\bibinfo  {journal} {SciPost Phys.}\ }\textbf {\bibinfo {volume} {12}},\ \bibinfo {pages} {166} (\bibinfo {year} {2022})}\BibitemShut {NoStop}%
\bibitem [{\citenamefont {Calandra~Buonaura}\ and\ \citenamefont {Sorella}(1998)}]{calandra1998}%
  \BibitemOpen
  \bibfield  {author} {\bibinfo {author} {\bibfnamefont {M.}~\bibnamefont {Calandra~Buonaura}}\ and\ \bibinfo {author} {\bibfnamefont {S.}~\bibnamefont {Sorella}},\ }\bibfield  {title} {\bibinfo {title} {Numerical study of the two-dimensional heisenberg model using a green function monte carlo technique with a fixed number of walkers},\ }\href {https://doi.org/10.1103/PhysRevB.57.11446} {\bibfield  {journal} {\bibinfo  {journal} {Phys. Rev. B}\ }\textbf {\bibinfo {volume} {57}},\ \bibinfo {pages} {11446} (\bibinfo {year} {1998})}\BibitemShut {NoStop}%
\bibitem [{\citenamefont {Sandvik}(1997)}]{sandvik1997}%
  \BibitemOpen
  \bibfield  {author} {\bibinfo {author} {\bibfnamefont {A.~W.}\ \bibnamefont {Sandvik}},\ }\bibfield  {title} {\bibinfo {title} {Finite-size scaling of the ground-state parameters of the two-dimensional heisenberg model},\ }\href {https://doi.org/10.1103/PhysRevB.56.11678} {\bibfield  {journal} {\bibinfo  {journal} {Phys. Rev. B}\ }\textbf {\bibinfo {volume} {56}},\ \bibinfo {pages} {11678} (\bibinfo {year} {1997})}\BibitemShut {NoStop}%
\bibitem [{\citenamefont {Hu}\ \emph {et~al.}(2013)\citenamefont {Hu}, \citenamefont {Becca}, \citenamefont {Parola},\ and\ \citenamefont {Sorella}}]{BeccaGutz2013}%
  \BibitemOpen
  \bibfield  {author} {\bibinfo {author} {\bibfnamefont {W.-J.}\ \bibnamefont {Hu}}, \bibinfo {author} {\bibfnamefont {F.}~\bibnamefont {Becca}}, \bibinfo {author} {\bibfnamefont {A.}~\bibnamefont {Parola}},\ and\ \bibinfo {author} {\bibfnamefont {S.}~\bibnamefont {Sorella}},\ }\bibfield  {title} {\bibinfo {title} {Direct evidence for a gapless ${Z}_{2}$ spin liquid by frustrating n\'eel antiferromagnetism},\ }\href {https://doi.org/10.1103/PhysRevB.88.060402} {\bibfield  {journal} {\bibinfo  {journal} {Phys. Rev. B}\ }\textbf {\bibinfo {volume} {88}},\ \bibinfo {pages} {060402} (\bibinfo {year} {2013})}\BibitemShut {NoStop}%
\bibitem [{\citenamefont {Ferrari}\ and\ \citenamefont {Becca}(2020)}]{ferrari2020}%
  \BibitemOpen
  \bibfield  {author} {\bibinfo {author} {\bibfnamefont {F.}~\bibnamefont {Ferrari}}\ and\ \bibinfo {author} {\bibfnamefont {F.}~\bibnamefont {Becca}},\ }\bibfield  {title} {\bibinfo {title} {Gapless spin liquid and valence-bond solid in the ${J}_{1}$-${J}_{2}$ heisenberg model on the square lattice: Insights from singlet and triplet excitations},\ }\href {https://doi.org/10.1103/PhysRevB.102.014417} {\bibfield  {journal} {\bibinfo  {journal} {Phys. Rev. B}\ }\textbf {\bibinfo {volume} {102}},\ \bibinfo {pages} {014417} (\bibinfo {year} {2020})}\BibitemShut {NoStop}%
\bibitem [{\citenamefont {Wang}\ \emph {et~al.}(2022)\citenamefont {Wang}, \citenamefont {Zhang},\ and\ \citenamefont {Sandvik}}]{wang2022}%
  \BibitemOpen
  \bibfield  {author} {\bibinfo {author} {\bibfnamefont {L.}~\bibnamefont {Wang}}, \bibinfo {author} {\bibfnamefont {Y.}~\bibnamefont {Zhang}},\ and\ \bibinfo {author} {\bibfnamefont {A.~W.}\ \bibnamefont {Sandvik}},\ }\bibfield  {title} {\bibinfo {title} {Quantum spin liquid phase in the shastry-sutherland model detected by an improved level spectroscopic method},\ }\href {https://doi.org/10.1088/0256-307X/39/7/077502} {\bibfield  {journal} {\bibinfo  {journal} {Chinese Physics Letters}\ }\textbf {\bibinfo {volume} {39}},\ \bibinfo {pages} {077502} (\bibinfo {year} {2022})}\BibitemShut {NoStop}%
\bibitem [{\citenamefont {Gong}\ \emph {et~al.}(2014)\citenamefont {Gong}, \citenamefont {Zhu}, \citenamefont {Sheng}, \citenamefont {Motrunich},\ and\ \citenamefont {Fisher}}]{gong2014}%
  \BibitemOpen
  \bibfield  {author} {\bibinfo {author} {\bibfnamefont {S.-S.}\ \bibnamefont {Gong}}, \bibinfo {author} {\bibfnamefont {W.}~\bibnamefont {Zhu}}, \bibinfo {author} {\bibfnamefont {D.~N.}\ \bibnamefont {Sheng}}, \bibinfo {author} {\bibfnamefont {O.~I.}\ \bibnamefont {Motrunich}},\ and\ \bibinfo {author} {\bibfnamefont {M.~P.~A.}\ \bibnamefont {Fisher}},\ }\bibfield  {title} {\bibinfo {title} {Plaquette ordered phase and quantum phase diagram in the spin-$\frac{1}{2}$ ${J}_{1}\text{\ensuremath{-}}{J}_{2}$ square heisenberg model},\ }\href {https://doi.org/10.1103/PhysRevLett.113.027201} {\bibfield  {journal} {\bibinfo  {journal} {Phys. Rev. Lett.}\ }\textbf {\bibinfo {volume} {113}},\ \bibinfo {pages} {027201} (\bibinfo {year} {2014})}\BibitemShut {NoStop}%
\bibitem [{\citenamefont {Wolf}\ \emph {et~al.}(2020)\citenamefont {Wolf}, \citenamefont {Debut}, \citenamefont {Sanh}, \citenamefont {Chaumond}, \citenamefont {Delangue}, \citenamefont {Moi}, \citenamefont {Cistac}, \citenamefont {Rault}, \citenamefont {Louf}, \citenamefont {Funtowicz}, \citenamefont {Davison}, \citenamefont {Shleifer}, \citenamefont {von Platen}, \citenamefont {Ma}, \citenamefont {Jernite}, \citenamefont {Plu}, \citenamefont {Xu}, \citenamefont {Scao}, \citenamefont {Gugger}, \citenamefont {Drame}, \citenamefont {Lhoest},\ and\ \citenamefont {Rush}}]{wolf2020}%
  \BibitemOpen
  \bibfield  {author} {\bibinfo {author} {\bibfnamefont {T.}~\bibnamefont {Wolf}}, \bibinfo {author} {\bibfnamefont {L.}~\bibnamefont {Debut}}, \bibinfo {author} {\bibfnamefont {V.}~\bibnamefont {Sanh}}, \bibinfo {author} {\bibfnamefont {J.}~\bibnamefont {Chaumond}}, \bibinfo {author} {\bibfnamefont {C.}~\bibnamefont {Delangue}}, \bibinfo {author} {\bibfnamefont {A.}~\bibnamefont {Moi}}, \bibinfo {author} {\bibfnamefont {P.}~\bibnamefont {Cistac}}, \bibinfo {author} {\bibfnamefont {T.}~\bibnamefont {Rault}}, \bibinfo {author} {\bibfnamefont {R.}~\bibnamefont {Louf}}, \bibinfo {author} {\bibfnamefont {M.}~\bibnamefont {Funtowicz}}, \bibinfo {author} {\bibfnamefont {J.}~\bibnamefont {Davison}}, \bibinfo {author} {\bibfnamefont {S.}~\bibnamefont {Shleifer}}, \bibinfo {author} {\bibfnamefont {P.}~\bibnamefont {von Platen}}, \bibinfo {author} {\bibfnamefont {C.}~\bibnamefont {Ma}}, \bibinfo {author} {\bibfnamefont {Y.}~\bibnamefont {Jernite}}, \bibinfo {author} {\bibfnamefont {J.}~\bibnamefont {Plu}}, \bibinfo
  {author} {\bibfnamefont {C.}~\bibnamefont {Xu}}, \bibinfo {author} {\bibfnamefont {T.~L.}\ \bibnamefont {Scao}}, \bibinfo {author} {\bibfnamefont {S.}~\bibnamefont {Gugger}}, \bibinfo {author} {\bibfnamefont {M.}~\bibnamefont {Drame}}, \bibinfo {author} {\bibfnamefont {Q.}~\bibnamefont {Lhoest}},\ and\ \bibinfo {author} {\bibfnamefont {A.~M.}\ \bibnamefont {Rush}},\ }\href@noop {} {\bibinfo {title} {Huggingface's transformers: State-of-the-art natural language processing}} (\bibinfo {year} {2020}),\ \Eprint {https://arxiv.org/abs/1910.03771} {arXiv:1910.03771 [cs.CL]} \BibitemShut {NoStop}%
\bibitem [{\citenamefont {Scherbela}\ \emph {et~al.}(2024)\citenamefont {Scherbela}, \citenamefont {Gerard},\ and\ \citenamefont {Grohs}}]{scherbela2024towards}%
  \BibitemOpen
  \bibfield  {author} {\bibinfo {author} {\bibfnamefont {M.}~\bibnamefont {Scherbela}}, \bibinfo {author} {\bibfnamefont {L.}~\bibnamefont {Gerard}},\ and\ \bibinfo {author} {\bibfnamefont {P.}~\bibnamefont {Grohs}},\ }\bibfield  {title} {\bibinfo {title} {Towards a transferable fermionic neural wavefunction for molecules},\ }\href {https://doi.org/https://doi.org/10.1038/s41467-023-44216-9} {\bibfield  {journal} {\bibinfo  {journal} {Nature Communications}\ }\textbf {\bibinfo {volume} {15}},\ \bibinfo {pages} {120} (\bibinfo {year} {2024})}\BibitemShut {NoStop}%
\bibitem [{\citenamefont {Cao}\ \emph {et~al.}(2023)\citenamefont {Cao}, \citenamefont {Pan}, \citenamefont {Wang},\ and\ \citenamefont {Zhang}}]{cao2023qecgpt}%
  \BibitemOpen
  \bibfield  {author} {\bibinfo {author} {\bibfnamefont {H.}~\bibnamefont {Cao}}, \bibinfo {author} {\bibfnamefont {F.}~\bibnamefont {Pan}}, \bibinfo {author} {\bibfnamefont {Y.}~\bibnamefont {Wang}},\ and\ \bibinfo {author} {\bibfnamefont {P.}~\bibnamefont {Zhang}},\ }\href@noop {} {\bibinfo {title} {qecgpt: decoding quantum error-correcting codes with generative pre-trained transformers}} (\bibinfo {year} {2023}),\ \Eprint {https://arxiv.org/abs/2307.09025} {arXiv:2307.09025 [quant-ph]} \BibitemShut {NoStop}%
\bibitem [{\citenamefont {Mendes-Santos}\ \emph {et~al.}(2021)\citenamefont {Mendes-Santos}, \citenamefont {Turkeshi}, \citenamefont {Dalmonte},\ and\ \citenamefont {Rodriguez}}]{santos2021}%
  \BibitemOpen
  \bibfield  {author} {\bibinfo {author} {\bibfnamefont {T.}~\bibnamefont {Mendes-Santos}}, \bibinfo {author} {\bibfnamefont {X.}~\bibnamefont {Turkeshi}}, \bibinfo {author} {\bibfnamefont {M.}~\bibnamefont {Dalmonte}},\ and\ \bibinfo {author} {\bibfnamefont {A.}~\bibnamefont {Rodriguez}},\ }\bibfield  {title} {\bibinfo {title} {Unsupervised learning universal critical behavior via the intrinsic dimension},\ }\href {https://doi.org/10.1103/PhysRevX.11.011040} {\bibfield  {journal} {\bibinfo  {journal} {Phys. Rev. X}\ }\textbf {\bibinfo {volume} {11}},\ \bibinfo {pages} {011040} (\bibinfo {year} {2021})}\BibitemShut {NoStop}%
\bibitem [{\citenamefont {van Nieuwenburg}\ \emph {et~al.}(2017)\citenamefont {van Nieuwenburg}, \citenamefont {Liu},\ and\ \citenamefont {Huber}}]{vannieuwenburg2017}%
  \BibitemOpen
  \bibfield  {author} {\bibinfo {author} {\bibfnamefont {E.~P.~L.}\ \bibnamefont {van Nieuwenburg}}, \bibinfo {author} {\bibfnamefont {Y.-H.}\ \bibnamefont {Liu}},\ and\ \bibinfo {author} {\bibfnamefont {S.~D.}\ \bibnamefont {Huber}},\ }\bibfield  {title} {\bibinfo {title} {Learning phase transitions by confusion},\ }\href {https://doi.org/10.1038/nphys4037} {\bibfield  {journal} {\bibinfo  {journal} {Nature Physics}\ }\textbf {\bibinfo {volume} {13}},\ \bibinfo {pages} {435–439} (\bibinfo {year} {2017})}\BibitemShut {NoStop}%
\bibitem [{\citenamefont {Zen}\ \emph {et~al.}(2020)\citenamefont {Zen}, \citenamefont {My}, \citenamefont {Tan}, \citenamefont {Hebert}, \citenamefont {Gattobigio}, \citenamefont {Miniatura}, \citenamefont {Poletti},\ and\ \citenamefont {Bressan}}]{zen2020}%
  \BibitemOpen
  \bibfield  {author} {\bibinfo {author} {\bibfnamefont {R.}~\bibnamefont {Zen}}, \bibinfo {author} {\bibfnamefont {L.}~\bibnamefont {My}}, \bibinfo {author} {\bibfnamefont {R.}~\bibnamefont {Tan}}, \bibinfo {author} {\bibfnamefont {F.}~\bibnamefont {Hebert}}, \bibinfo {author} {\bibfnamefont {M.}~\bibnamefont {Gattobigio}}, \bibinfo {author} {\bibfnamefont {C.}~\bibnamefont {Miniatura}}, \bibinfo {author} {\bibfnamefont {D.}~\bibnamefont {Poletti}},\ and\ \bibinfo {author} {\bibfnamefont {S.}~\bibnamefont {Bressan}},\ }\href@noop {} {\bibinfo {title} {Finding quantum critical points with neural-network quantum states}} (\bibinfo {year} {2020}),\ \Eprint {https://arxiv.org/abs/2002.02618} {arXiv:2002.02618 [physics.comp-ph]} \BibitemShut {NoStop}%
\bibitem [{\citenamefont {Lewis}\ \emph {et~al.}(2023)\citenamefont {Lewis}, \citenamefont {Huang}, \citenamefont {Tran}, \citenamefont {Lehner}, \citenamefont {Kueng},\ and\ \citenamefont {Preskill}}]{lewis2023}%
  \BibitemOpen
  \bibfield  {author} {\bibinfo {author} {\bibfnamefont {L.}~\bibnamefont {Lewis}}, \bibinfo {author} {\bibfnamefont {H.-Y.}\ \bibnamefont {Huang}}, \bibinfo {author} {\bibfnamefont {V.~T.}\ \bibnamefont {Tran}}, \bibinfo {author} {\bibfnamefont {S.}~\bibnamefont {Lehner}}, \bibinfo {author} {\bibfnamefont {R.}~\bibnamefont {Kueng}},\ and\ \bibinfo {author} {\bibfnamefont {J.}~\bibnamefont {Preskill}},\ }\href@noop {} {\bibinfo {title} {Improved machine learning algorithm for predicting ground state properties}} (\bibinfo {year} {2023}),\ \Eprint {https://arxiv.org/abs/2301.13169} {arXiv:2301.13169 [quant-ph]} \BibitemShut {NoStop}%
\bibitem [{\citenamefont {Schmitt}\ and\ \citenamefont {Heyl}(2020)}]{SchmittDyn2020}%
  \BibitemOpen
  \bibfield  {author} {\bibinfo {author} {\bibfnamefont {M.}~\bibnamefont {Schmitt}}\ and\ \bibinfo {author} {\bibfnamefont {M.}~\bibnamefont {Heyl}},\ }\bibfield  {title} {\bibinfo {title} {Quantum many-body dynamics in two dimensions with artificial neural networks},\ }\href {https://doi.org/10.1103/PhysRevLett.125.100503} {\bibfield  {journal} {\bibinfo  {journal} {Phys. Rev. Lett.}\ }\textbf {\bibinfo {volume} {125}},\ \bibinfo {pages} {100503} (\bibinfo {year} {2020})}\BibitemShut {NoStop}%
\bibitem [{\citenamefont {Mendes-Santos}\ \emph {et~al.}(2023)\citenamefont {Mendes-Santos}, \citenamefont {Schmitt},\ and\ \citenamefont {Heyl}}]{HeylDyn2023}%
  \BibitemOpen
  \bibfield  {author} {\bibinfo {author} {\bibfnamefont {T.}~\bibnamefont {Mendes-Santos}}, \bibinfo {author} {\bibfnamefont {M.}~\bibnamefont {Schmitt}},\ and\ \bibinfo {author} {\bibfnamefont {M.}~\bibnamefont {Heyl}},\ }\bibfield  {title} {\bibinfo {title} {Highly resolved spectral functions of two-dimensional systems with neural quantum states},\ }\href {https://doi.org/10.1103/PhysRevLett.131.046501} {\bibfield  {journal} {\bibinfo  {journal} {Phys. Rev. Lett.}\ }\textbf {\bibinfo {volume} {131}},\ \bibinfo {pages} {046501} (\bibinfo {year} {2023})}\BibitemShut {NoStop}%
\bibitem [{\citenamefont {Fishman}\ \emph {et~al.}(2022)\citenamefont {Fishman}, \citenamefont {White},\ and\ \citenamefont {Stoudenmire}}]{itensor}%
  \BibitemOpen
  \bibfield  {author} {\bibinfo {author} {\bibfnamefont {M.}~\bibnamefont {Fishman}}, \bibinfo {author} {\bibfnamefont {S.}~\bibnamefont {White}},\ and\ \bibinfo {author} {\bibfnamefont {M.}~\bibnamefont {Stoudenmire}},\ }\bibfield  {title} {\bibinfo {title} {{The ITensor Software Library for Tensor Network Calculations}},\ }\href {https://doi.org/10.21468/SciPostPhysCodeb.4} {\bibfield  {journal} {\bibinfo  {journal} {SciPost Phys. Codebases}\ ,\ \bibinfo {pages} {4}} (\bibinfo {year} {2022})}\BibitemShut {NoStop}%
\bibitem [{\citenamefont {Novak}\ \emph {et~al.}(2022)\citenamefont {Novak}, \citenamefont {Sohl-Dickstein},\ and\ \citenamefont {Schoenholz}}]{novak22}%
  \BibitemOpen
  \bibfield  {author} {\bibinfo {author} {\bibfnamefont {R.}~\bibnamefont {Novak}}, \bibinfo {author} {\bibfnamefont {J.}~\bibnamefont {Sohl-Dickstein}},\ and\ \bibinfo {author} {\bibfnamefont {S.~S.}\ \bibnamefont {Schoenholz}},\ }\bibfield  {title} {\bibinfo {title} {Fast finite width neural tangent kernel},\ }in\ \href {https://proceedings.mlr.press/v162/novak22a.html} {\emph {\bibinfo {booktitle} {Proceedings of the 39th International Conference on Machine Learning}}},\ \bibinfo {series} {Proceedings of Machine Learning Research}, Vol.\ \bibinfo {volume} {162}\ (\bibinfo  {publisher} {PMLR},\ \bibinfo {year} {2022})\ pp.\ \bibinfo {pages} {17018--17044}\BibitemShut {NoStop}%
\bibitem [{\citenamefont {McInnes}\ \emph {et~al.}(2020)\citenamefont {McInnes}, \citenamefont {Healy},\ and\ \citenamefont {Melville}}]{mcinnes2020}%
  \BibitemOpen
  \bibfield  {author} {\bibinfo {author} {\bibfnamefont {L.}~\bibnamefont {McInnes}}, \bibinfo {author} {\bibfnamefont {J.}~\bibnamefont {Healy}},\ and\ \bibinfo {author} {\bibfnamefont {J.}~\bibnamefont {Melville}},\ }\href@noop {} {\bibinfo {title} {Umap: Uniform manifold approximation and projection for dimension reduction}} (\bibinfo {year} {2020}),\ \Eprint {https://arxiv.org/abs/1802.03426} {arXiv:1802.03426 [stat.ML]} \BibitemShut {NoStop}%
\end{thebibliography}%

\end{document}